\documentclass[showpacs,preprintnumbers,pre]{revtex4}
\usepackage{graphicx}
\usepackage{bm}
\begin{document}
\title{
Microstructure and Velocity of Field-Driven SOS Interfaces:
Analytic Approximations and Numerical Results 
}
\author{
Per Arne Rikvold$^{1,2,}$}\email{rikvold@csit.fsu.edu} 
\author{M.~Kolesik$^{3,4,}$}\email{kolesik@acms.arizona.edu}
\affiliation{
$^1$Center for Materials Research and Technology, 
School of Computational Science and Information Technology, and 
Department of Physics,
Florida State University, Tallahassee, Florida 32306-4350\\
$^2$Department of Fundamental Sciences, Faculty of Integrated Human Studies, 
Kyoto University, Kyoto 606, Japan\\
$^3$Institute of Physics, Slovak Academy of Sciences,
Bratislava, Slovak Republic\\
$^4$Optical Sciences Center, University of Arizona,
Tucson, Arizona 85721
}
\date{\today}

\begin{abstract}
The local structure of a solid-on-solid (SOS) interface in a two-dimensional 
kinetic Ising ferromagnet with single-spin-flip Glauber dynamics, which is 
driven far from equilibrium by an applied field, is studied by an analytic 
mean-field, nonlinear-response theory 
[P.~A.\ Rikvold and M.~Kolesik, J.\ Stat.\ Phys.\ {\bf 100}, 377 (2000)] 
and by dynamic Monte Carlo simulations. The probability density of the height 
of an individual step in the surface is obtained, both analytically and by 
simulation. The width of the probability density is found to increase 
dramatically 
with the magnitude of the applied field, with close agreement between the 
theoretical predictions and the simulation results. Excellent agreement between 
theory and simulations is also found for the field-dependence and 
anisotropy of the interface velocity. The 
joint distribution of nearest-neighbor step heights is obtained by simulation. 
It shows increasing correlations with increasing field, similar to the skewness 
observed in other examples of growing surfaces. 
\end{abstract}

\pacs{ 
68.35.Ct 
75.60.Jk 
68.43.Jk 
05.10.Ln 
}

\maketitle


\section{Introduction}
\label{sec:INTRO}

The motion of surfaces and interfaces plays a central role in many 
scientific and technological disciplines. In particular, the dynamics of 
interfaces such as phase- and grain boundaries in solid materials \cite{MEND00}
and domain walls in magnets \cite{RIKVturkey} and ferroelectrics 
heavily influence both dynamic and static materials properties. Among 
interfaces characteristic of two-dimensional systems are steps on 
crystal surfaces \cite{Surfstep}, 
domain walls in thin magnetic and dielectric films \cite{RIKVturkey}, 
and boundaries between different types of vegetation such as savanna and 
rainforest \cite{RAINF}. 

An enormous amount of work in recent years has been devoted to the dynamics and 
structure of moving and growing interfaces \cite{BARA95,MEAK98}. However, 
despite the fact that many important interface properties, such as mobility 
and catalytic and chemical 
activity, are largely determined by the {\it microscopic\/} interface 
structure, most of this effort has been concentrated on large-scale scaling 
properties. 
Although the detailed atomistic mechanisms by which interfaces move are 
often not known, useful understanding can be obtained from stochastic models 
in which the motion occurs through random nucleation and migration of 
local topological features such as kinks and steps \cite{MEND00}. 
It is therefore important to gain better insight 
into how, for different stochastic  dynamics, the driving force (such as an 
applied magnetic or electric field, a chemical-potential gradient, 
or the amount of rainfall in the case of models of vegetation distribution) 
may alter 
the microscopic structure of the interface, thereby leading to a highly 
nonlinear velocity response. 

In a previous paper \cite{RIKV00} we introduced 
a dynamic mean-field approximation for the microstructure 
of an interface in a two-dimensional kinetic Ising ferromagnet 
with a single-spin-flip Glauber dynamic \cite{KAWA72,LAND00}, 
driven by an applied field \cite{DEVI92,SPOH93}. 
This model is directly applicable to many magnetic and 
ferroelectric systems and other cases where the interface dynamics are not 
inhibited by coupling to a conserved field \cite{RAMO99,NOVO99B}. 
Based on the resulting local interface structure, we obtained a 
nonlinear-response approximation for the steady-state propagation velocity, 
which was shown to be in good agreement with dynamic Monte Carlo (MC)
simulations for a wide range of fields and temperatures. However, since the 
approximation was based on the Burton-Cabrera-Frank (BCF) 
unrestricted solid-on-solid (SOS) 
model \cite{BURT51}, the overhangs and bubbles in the Ising interface were 
handled in an uncontrolled way. Here we therefore 
consider the performance of our 
approximation for the unrestricted SOS model, 
so that overhangs and bubbles are absent at all times by definition of 
the model. In particular, we obtain the surface velocity under the Glauber 
dynamic as a function of 
applied field and temperature, as well as its anisotropy for 
tilt angles between $0^\circ$ and $45^\circ$. 

In a recent paper \cite{RIKV01} we showed that the microscopic interface 
structure, and thus the mobility, can depend dramatically on the details of the 
dynamics. The most significant difference is between dynamics in which the 
transition probabilities of the local spins factorize into one part that 
depends only on the change in interaction energy due to the transition and 
one that depends only on the change in the field energy 
(soft dynamics \cite{MARR99}), and dynamics that cannot be factorized 
in this way (hard dynamics \cite{MARR99}). 
For soft dynamics, the interface structure for all values of the field 
remains the {\it same\/} as in equilibrium at zero field \cite{RIKV01}, 
so that the mobility can be calculated exactly at the level of 
linear response. In the present paper we concentrate on the standard Glauber 
dynamic \cite{KAWA72,LAND00}, 
which belongs to the class of hard dynamics and thus leads to 
more complicated and interesting behavior. 

Both the driven Ising and SOS surfaces belong to the Kardar-Parisi-Zhang 
(KPZ) dynamic universality class \cite{BARA95,KARD86}, in which 
the macroscopic, stationary distribution for flat,
moving interfaces is Gaussian, corresponding to a random walk with independent
increments. Nevertheless, the step heights in several discrete models in
this class are correlated at {\it short\/} distances \cite{NEER97,KORN00}. 
In the mean-field approximation 
developed here, these short-range correlations are 
ignored. The resulting discrepancies, which are minor, are elucidated 
by comparison with MC simulations. 

The remainder of this paper is organized as follows. 
In Sec.~\ref{sec:MODEL} we introduce the SOS interface model and derive a 
linear-response approximation for its velocity in a nonzero external field. 
In Sec.~\ref{sec:NLR} we develop a mean-field approximation for the time 
evolution of the single-step probability density function (pdf), 
as well as for  its stationary form. 
The latter enables us to extend the approximation for the interface 
structure and velocity to a nonlinear-response level. 
The analytical approximations are compared with dynamic MC simulations 
in Sec.~\ref{sec:MC}. 
In Sec.~\ref{sec:MC0} we numerically solve the mean-field equation 
of motion for the single-step 
pdf and compare the resulting values of the time-dependent 
average interface step height with MC simulations. 
In Sec.~\ref{sec:MCss} we compare the simulated stationary 
single-step pdfs with the theoretical predictions. 
In Sec.~\ref{sec:MCi} we compare the simulated stationary 
interface velocity with the 
theoretical predictions for various values of applied field, temperature, and 
interface tilt angle. 
In Sec.~\ref{sec:MCc} we compare simulations and analytical predictions 
for the detailed stationary interface structure, including the asymmetry
of the simulated interface at nonzero fields. 
A summary and conclusions are found in Sec.~\ref{sec:DISC}.

\section{Model and Dynamics}
\label{sec:MODEL}

The original BCF SOS model considers an interface in a lattice gas or $S=1/2$ 
Ising system on a square lattice of unit lattice constant 
as a series of integer-valued steps 
$\delta(x)$, parallel to the $y$ axis. The interface height is thus a 
single-valued integer 
function of the $x$-coordinate, with steps at integer values of $x$. 
The configuration 
energy is given by the nearest-neighbor Ising Hamiltonian with anisotropic, 
ferromagnetic interactions $J_x$ and $J_y$ in the $x$ and $y$ direction, 
respectively: 
\begin{equation}
{\cal H} = -\sum_{x,y} s_{x,y} \left( J_x s_{x+1,y} + J_y s_{x,y+1} 
+ H \right) 
\;. 
\label{eq:ham}
\end{equation}
Here the two states at lattice site $(x,y)$ are denoted by $s_{x,y} = \pm 1$, 
and $\sum_{x,y}$ runs over all sites. The quantity 
$H$ is the applied field, and the interface is introduced by fixing 
$s_{x,y}=+1$ and $-$1 for large negative and positive $y$, respectively.  
Without loss of generality we take $H \ge 0$, such 
that the interface on average moves in the positive $y$ direction. 

A single-spin-flip 
(nonconservative) dynamic which satisfies detailed balance, such as the 
Metropolis or Glauber algorithms \cite{KAWA72,LAND00}, ensures the approach to 
equilibrium, which in this case is a 
uniformly positive phase with the interface pushed off to positive infinity. 
Such algorithms are defined by a single-spin transition probability, 
$W(s_{x,y} \rightarrow -s_{x,y}) = W(\beta \Delta E , \beta U)$. Here 
$\beta$ is the inverse of the temperature $T$ (Boltzmann's constant is 
taken as unity), $\Delta E$ is the energy change corresponding to a
successful spin flip, and the optional parameter $U$ is an energy barrier 
between the two states that enters into Arrhenius-type stochastic dynamics 
\cite{KANG89} and other dynamics that include a transition state 
\cite{TDA}. The detailed-balance condition is expressed as 
$W(\beta \Delta E, \beta U) / W(- \beta \Delta E , \beta U) 
= e^{- \beta \Delta E}$, where the right-hand side is independent of $U$. 
(In the case of soft dynamics, the detailed-balance condition is satisfied 
independently for the two parts of $W$.) 
In order to preserve the
SOS configuration at all times, flips are allowed only at 
sites which have exactly one broken bond in the $y$ direction. 

With the Ising Hamiltonian there are only a finite 
number of different values of $\Delta E$, and the spins can therefore be 
divided into classes \cite{SPOH93,BORT75,NOVO95A}, 
labeled by the spin value $s$ and the number 
of broken bonds between the spin and its nearest neighbors in the $x$ and 
$y$ direction, $j$ and $k$, respectively. 
The ten spin classes consistent with the SOS model are denoted $jks$ 
with $j \in \{0,1,2\}$ and $k \in \{0,1\}$. They are listed in 
Table~\ref{table:class}. 

In this paper, as in Ref.~\cite{RIKV00}, we use the standard discrete-time 
Glauber dynamic with the transition probability 
\cite{KAWA72,LAND00} 
\begin{equation}
W_{\rm G}(s_{x,y} \rightarrow -s_{x,y}) 
= \left[ 1 + e^{ \beta \Delta E } \right]^{-1} 
.
\label{eq:glau}
\end{equation}
Time is measured in MC steps per spin (MCSS). 

In the BCF SOS model the heights of the individual 
steps are assumed to be statistically independent and 
identically distributed. The step-height probability density function (pdf) 
is given by the interaction energy corresponding to the $|\delta(x)|$ broken 
$J_x$-bonds between spins in the columns centered 
at $(x-1/2)$ and  $(x+1/2)$ as  
\begin{equation}
p[\delta(x)] = Z(\phi)^{-1} X^{|\delta(x)|}
\ e^{ \gamma(\phi) \delta(x) } \;. 
\label{eq:step_pdf}
\end{equation}
The Boltzmann factor $X = e^{- 2 \beta J_x}$ determines the width of the pdf, 
and $\gamma(\phi)$ is a Lagrange multiplier which maintains the mean step 
height at an $x$-independent value, $\langle \delta(x) \rangle = \tan \phi$, 
where $\phi$ is the average angle between the interface and the $x$ axis. 
The partition function is 
\begin{equation}
Z(\phi)
=
\sum_{\delta = -\infty}^{+\infty} X^{|\delta|} e^{ \gamma(\phi) \delta } 
= 
\frac{1-X^2}{1 - 2 X \cosh \gamma(\phi) + X^2} 
\label{eq:Z}
\end{equation}
with $\gamma(\phi)$ given by 
\begin{equation}
e^{\gamma (\phi)} 
= 
\frac{ \left(1+X^2 \right)\tan \phi 
+ \left[ \left( 1 - X^2 \right)^2 \tan^2 \phi + 4 X^2 \right]^{1/2}}
{2 X \left( 1 + \tan \phi \right)} 
\label{eq:chgam}
\end{equation}
(see details in Ref.~\cite{RIKV00}). 
For $\phi = 0$, $\gamma(\phi) = 0$, and the partition function simplifies to 
\begin{equation}
Z(0) = (1+X)/(1-X) \;.
\label{eq:Z0}
\end{equation} 

The mean spin-class 
populations, $\langle n(jks) \rangle$, are all obtained from the 
product of the independent pdfs for $\delta(x)$ and $\delta(x$+1). 
Symmetry of $p[\delta(x)]$ under the transformation 
$(x,\phi,\delta) \rightarrow (-x,-\phi,-\delta)$ ensures that 
$\langle n(jk-) \rangle = \langle n(jk+) \rangle$ for all $j$ and $k$. 
Numerical results illustrating the breakdown of this up/down symmetry for 
large $|H|$ are discussed in Sec.~\ref{sec:MCc}. 
As discussed in Ref.~\cite{RIKV00}, calculation of the individual class 
populations 
is straightforward but somewhat tedious, especially for nonzero $\phi$. 
The final results are summarized in Table~\ref{table:class2}. 

Whenever a spin  flips from $-$1 to +1, 
the corresponding column of the interface advances by one lattice constant 
in the $y$ direction. Conversely, the column 
recedes by one lattice constant when a spin 
flips from +1 to $-$1. The corresponding energy changes are 
given in the third column in Table~\ref{table:class}. 
Since the spin-class populations on both sides of the 
interface are equal in this approximation, the contribution 
to the mean velocity in the $y$ direction 
from sites in the classes $jk-$ and $jk+$ becomes 
\begin{equation}
\langle v_y(jk) \rangle 
= 
W \left( \beta \Delta E(jk-) ,\beta U  \right)
-
W \left( \beta \Delta E(jk+) , \beta U \right) 
 \;. 
\label{eq_generalv}
\end{equation}
The results corresponding to 
the hard Glauber transition probabilities used here, Eq.~(\ref{eq:glau}), 
are listed in the last column of Table~\ref{table:class2}.  
The mean propagation velocity perpendicular to the interface becomes  
\begin{equation}
\langle v_\perp (T,H,\phi) \rangle 
= 
\cos \phi \sum_{j,k} \langle n(jks) \rangle \langle v_y (jk) \rangle 
\;, 
\label{eq:totalv}
\end{equation}
where the sum runs over the classes included in the tables. 
While the general result is cumbersome 
if written out in detail, the special case of $\phi=0$ leads to a 
compact formula: 
\begin{eqnarray}
\langle v_\perp (T,H,0) \rangle 
&=&
\frac{\tanh (\beta H)}{(1+X)^2} 
\left\{
2 X 
+ 
\frac{1+X^2}
{1 + \left[\frac{\sinh (2 \beta J_x)}{\cosh (\beta H)}\right]^2} 
\right\} 
\;.  
\label{eq:totalv0}
\end{eqnarray}
It was shown in Ref.~\cite{RIKV00} 
that Eq.~(\ref{eq:totalv}) reduces to the results for 
the single-step \cite{DEVI92,SPOH93,MEAK86,PLIS87} and the polynuclear growth 
\cite{DEVI92,KRUG89,KERT89}
models at low temperatures for strong and weak fields, respectively. 

\section{Nonlinear-Response} 
\label{sec:NLR} 

With $X = e^{-2 \beta J_x}$, the results in Table~\ref{table:class2} 
correspond to a 
linear-response approximation. In Ref.~\cite{RIKV00} we developed a mean-field 
approximation leading to a field-dependent $X(T,H)$, based on a 
detailed-balance argument for the stationary state. 
Here we show that this detailed-balance relation follows naturally from a 
dynamic mean-field approximation for the equation of motion for the 
single-step pdf during the approach to the stationary state. 

We denote the total transition probability for the height of the single step 
at $x$ to change from $\delta(x)$ to $\delta(x) \pm 1$
as ${\cal W}[\delta(x) \rightarrow \delta(x) \pm1]$. 
In terms of these transition probabilities, the equation of motion for the 
single-step pdf, $p[\delta(x), t]$ becomes
\begin{eqnarray}
\frac{{d} p[\delta(x), t]}{{d} t} 
&=&
- p[\delta(x),t] \{{\cal W}[\delta(x) \rightarrow \delta(x) - 1] 
\nonumber\\
&& \;\; + {\cal W}[\delta(x) \rightarrow \delta(x) + 1] \} \nonumber\\
&&
+ p[\delta(x) + 1] {\cal W}[\delta(x) +1 \rightarrow \delta(x)] \nonumber\\
&&
+ p[\delta(x) - 1] {\cal W}[\delta(x) -1 \rightarrow \delta(x)]
\;,
\label{eq:eom}
\end{eqnarray}
where the coupling to the joint multistep 
probability density is hidden in the single-step transition rates $\cal W$. 

\begin{figure}[ht] 
\includegraphics[angle=270,width=.48\textwidth]{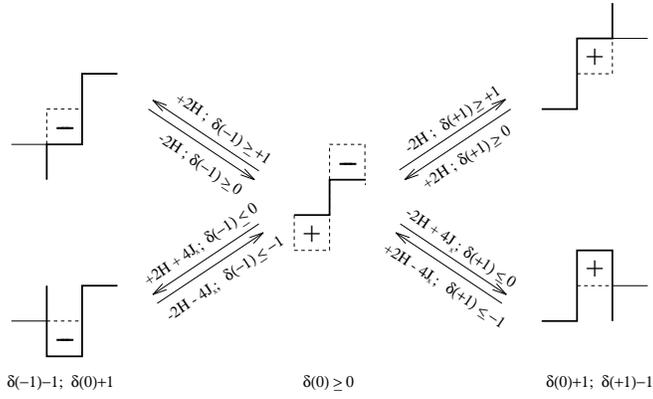} \\
\caption[]{
Figure for calculating the single-step transition rates in 
Eq.~(\protect\ref{eq:rates}). Interface configurations are shown by bold 
line segments. Spins above (in front of) 
the interface equal $-$1, and spins below (behind) the interface equal +1. 
At the center is shown a step $\delta(0) \ge 1$ (here shown as 
$\delta(0) = +1$). A transition to $\delta(0) + 1$ can be effected by flipping 
either of the two spins in the dashed boxes. The resulting configurations, 
which depend on the heights of the neighboring steps, are shown to the right and
left in the figure. The corresponding energy changes and conditions on the 
neighboring step heights are given next to the arrows. The arrows pointing 
outward from the center of the figure correspond to the transition described 
above, while the arrows pointing toward the center correspond to the reverse 
transition, $\delta(0) + 1 \rightarrow \delta(0)$. 
After Ref.~\protect\cite{RIKV00}. 
}
\label{fig:DE}
\end{figure}

To obtain an approximation for ${\cal W}[\delta \rightarrow \delta \pm 1]$, 
we employ the same mean-field assumption of independent steps 
as in equilibrium. For $\delta(x) \ge 1$ to increase to $\delta(x)+1$, 
either the 
spin in front of the interface at $x$+1/2 can flip from $-$1 to +1, or the 
spin behind the interface at $x$$-$1/2 
can flip from +1 to $-$1. In each of these 
cases, $\Delta E$ can have two different values, depending 
on the value of $\delta(x$+1) and $\delta(x$$-$1), respectively. The same 
argument holds also for the reverse transition, 
$\delta(x)+1 \rightarrow \delta(x)$. The energy changes and corresponding 
conditions on $\delta(x$+1) and $\delta(x$$-$1), which 
are shown in Fig.~\ref{fig:DE}, yield 
\begin{eqnarray}
{\cal W}[\delta \rightarrow \delta \pm 1 , t] 
&=& 
\frac{1}{2} \Big\{ 
[W(-2H) + W(+2H)] \Pi_\pm(t) \nonumber\\
&& + [W(-2H+4J_x) + W(+2H+4J_x) ] [1-\Pi_\pm(t)] 
\Big\}
\nonumber\\
{\cal W}[\delta \pm 1 \rightarrow \delta, t] 
&=& 
\frac{1}{2} \Big\{ 
[W(-2H) + W(+2H)] [1- \Pi_\mp(t)] \nonumber\\
&& + [W(-2H-4J_x) + W(+2H-4J_x)] \Pi_\mp(t)  
\Big\}
\;,
\label{eq:rates}
\end{eqnarray}
where the upper signs refer to $\delta \ge +1$ and the lower signs to 
$\delta \le -1$, $\Pi_+(t) = \sum_{\delta = +1}^{+\infty} p[\delta,t]$, and 
$\Pi_-(t)= \sum_{\delta = -1}^{-\infty} p[\delta,t]$. For simplicity, we 
here write the single-site transition rates $W(\beta \Delta E , \beta U)$ 
as $W(\Delta E)$.

In the stationary limit, Eqs.~(\ref{eq:eom}) 
and~(\ref{eq:step_pdf}) lead to the detailed-balance condition 
\begin{equation}
X(T,H) e^{\pm \gamma(\phi)} 
\equiv 
\frac{p[\delta \pm 1]}{p[\delta]} 
=
\frac{{\cal W}[\delta \rightarrow \delta \pm 1]}
{{\cal W}[\delta \pm 1 \rightarrow \delta]}
\label{eq:step_1}\;, 
\end{equation} 
where the upper and lower signs have the same interpretation as in 
Eq.~(\ref{eq:rates}). Using Eqs.~(\ref{eq:step_pdf}) and 
(\ref{eq:Z}), we get the stationary values for $\Pi_+$ and $\Pi_-$,  
\begin{eqnarray}
\Pi_+ 
&=& 
\frac{X e^{\gamma(\phi)} \left( 1 - X e^{-\gamma(\phi)} \right)}{1 - X^2} 
\nonumber\\
\Pi_- 
&=& 
\frac{X e^{-\gamma(\phi)} \left( 1 - X e^{\gamma(\phi)} \right)}{1 - X^2} 
\;,
\label{eq:pieq}
\end{eqnarray}
which, when inserted together with 
Eq.~(\ref{eq:rates}) in Eq.~(\ref{eq:step_1}), yield a self-consistency 
equation for $X$. The self-consistency equation reduces to a linear equation 
for $X^2$, and with the help of the detailed-balance condition for the 
spin transition rates $W$, the solution takes the form 
\begin{equation}
X(T,H) = e^{-2 \beta J_x} 
\left\{
\frac{e^{-2 \beta H}W(-2H-4J_x) + e^{2\beta H}W(+2H-4J_x)}
{W(-2H-4J_x) + W(+2H-4J_x)}
\right\}^{1/2}
\;,
\label{eq:XTH}
\end{equation}
which is independent of $\gamma(\phi)$. (This solution is the same as 
the one obtained in 
Ref.~\cite{RIKV00}, but the derivation given in that paper did not 
explicitly show that all dependence on $\gamma(\phi)$ cancels out.) 

Equation~(\ref{eq:XTH}) shows that $X(T,H)$ depends on the specific 
dynamic, except for $H=0$, where it reduces to its equilibrium value, 
$X(T,0) = e^{-2 \beta J_x}$, for all dynamics that satisfy detailed balance. 
A situation in which the $H$-dependence in Eq.~(\ref{eq:XTH})
cancels out, is that of the soft dynamics discussed in Sec.~\ref{sec:INTRO}. 
(The barrier energy $U$ can be contained in one or the other of the factors.) 
In Ref.~\cite{RIKV01} we demonstrated that a soft dynamic yields an SOS interface 
that is identical to the equilibrium SOS interface at $H=0$ and the same 
temperature, regardless of the value of $H$. 
Neither the Glauber dynamic used here,   
nor the equally common Metropolis dynamic with transition probability  
$W_{\rm M}(s_{x,y} \rightarrow -s_{x,y}) 
= \min \left[ 1 , e^{ - \beta \Delta E } \right]$ \cite{KAWA72,LAND00}, 
satisfies this factorization condition. Such hard dynamics lead to a
nontrivial field dependence in $X$. 
Inserting the Glauber dynamic defined by Eq.~(\ref{eq:glau}) 
into Eq.~(\ref{eq:XTH}), we explicitly get 
\begin{equation}
X_{\rm G}(T,H) = 
e^{-2 \beta J_x} 
\left\{
\frac{e^{2 \beta J_x} \cosh(2 \beta H) + e^{-2\beta J_x}}
{e^{-2 \beta J_x} \cosh(2 \beta H) + e^{2\beta J_x}}
\right\}^{1/2}
\;.
\label{eq:XG}
\end{equation}

All the results for the spin-class populations of the 
zero-field equilibrium interface, which are listed in Table~\ref{table:class2}, 
can now be applied to obtain a nonlinear-response approximation for the 
steady-state propagation velocity of flat, driven interfaces with hard dynamics. 
This simply requires replacing the zero-field $X=e^{-2 \beta J_x}$ used in the 
linear-response approximation by the 
field-dependent $X(T,H)$, obtained from Eq.~(\ref{eq:XTH}) using the 
transition probabilities corresponding to the particular dynamic used. 
For soft dynamics the linear-response result with $X=e^{-2 \beta J_x}$ is exact 
\cite{RIKV01}. 

A physical reason for the marked difference between hard and soft dynamics 
is best seen by comparing concrete examples of dynamics in the two classes, 
such as the hard Glauber dynamics used here, Eq.~(\ref{eq:glau}), and the 
soft Glauber dynamics used in Ref.~\cite{RIKV01}, 
\begin{equation}
W_{\rm SG}(s_{x,y} \rightarrow -s_{x,y}) 
= 
\left[ 1 + e^{\beta \Delta E_H } \right]^{-1}  
\cdot
\left[ 1 + e^{\beta \Delta E_J } \right]^{-1}  
\;, 
\label{eq:SG}
\end{equation}
in the case of a very strong field. In the hard case, the effect of the field 
completely dominates the transition rates, such that the rate is near unity 
for transitions that bring a spin parallel to the applied field, and near 
zero for transitions in the opposite direction, {\it irrespective of the 
change in interaction energy\/}. In the soft case, the probability of 
bringing a spin antiparallel to the field is also near zero, but 
the probabilities of different 
transitions bringing a spin parallel to the field differ according to the 
corresponding change in the interaction energy, as given by the second factor in 
Eq.~(\ref{eq:SG}). 

In the next Section we show that this nonlinear-response approximation 
gives very good agreement with MC simulations of driven, flat SOS interfaces 
evolving under the hard Glauber dynamic 
for a wide range of fields and temperatures. 

\section{Comparison with Monte Carlo Simulations} 
\label{sec:MC} 

We have compared the analytical estimates of step-height distributions, 
propagation velocities, 
and spin-class populations developed above with MC simulations of the 
same model for $J_x = J_y = J$. The details of our particular implementation 
of the discrete-time $n$-fold way rejection-free MC 
algorithm \cite{BORT75} are the same as described in Ref.~\cite{RIKV00}, 
except that only transitions from the classes with one broken $y$-bond 
($k=1$) are allowed. 
By keeping only the interface sites in memory, the algorithm is not subject 
to any size restriction in the $y$ direction, and simulations can be carried out 
for arbitrarily long times. 

The numerical results presented here are based on MC simulations at the two 
temperatures, $T = 0.2T_c$ and~0.6$T_c$ ($T_c = -2J/\ln(\sqrt{2}-1)$ 
is the critical temperature for the isotropic, 
square-lattice Ising model \cite{ONSA44}), with  
$L_x = 10\,000$ and fixed $\phi$ between 0 and $45^\circ$. 
In order to ensure stationarity we ran the simulation for 5\,000 
$n$-fold way updates per updatable spin (UPS)  
before taking any measurements (100\,000~UPS for some of the strongest fields 
and largest values of $\phi$ at $T=0.2T_c$).  Exploratory simulations with both 
larger and smaller $L_x$ (up to 100\,000) and 
``warm-up'' times (see Sec.~\ref{sec:MC0}) showed that the values used in the 
production runs were sufficient to ensure a stationary 
interface.  Stationary 
class populations and interface velocities were averaged over 50\,000~UPS. 
In the stationary limit 1~UPS corresponds to between 2~MCSS for strong fields at 
both temperatures, and about 75~MCSS for $H=0$ at $T = 0.2T_c$. 
Adequate statistics for one- and two-step pdfs was ensured by the large $L_x$, 
ten times the value used in Ref.~\cite{RIKV00}. 
Each data point took approximately nine hours on a DEC Alpha Unix workstation
or four hours on a Pentium II PC running Linux. 

\subsection{Approach to stationarity}
\label{sec:MC0}

In order both to check the applicability of the mean-field approximation at 
early times, and to decide the approximate time needed to reach the stationary 
state, we first studied the transient behavior of the 
average step height $\langle | \delta | \rangle$ for $\phi = 0$ at 
$T = 0.2J$ and 0.6$J$. 

In the dynamic MC simulation $\langle | \delta | \rangle$ was 
measured during a ``time window'' which was opened 
after a specified number of UPS, corresponding to 
a given approximate average 
evolution time $t$ in MCSS. To obtain an optimum balance between time resolution 
and accuracy, the width of the window was varied from approximately 
one MCSS at early times, to about ten MCSS at late times, and five independent 
runs were performed for each value of $t$. Standard errors for $t$ and 
$\langle | \delta | \rangle$ were estimated in the usual way 
from the spread in their measured values over the five realizations. 

\begin{figure}[ht] 
\includegraphics[angle=270,width=.48\textwidth]{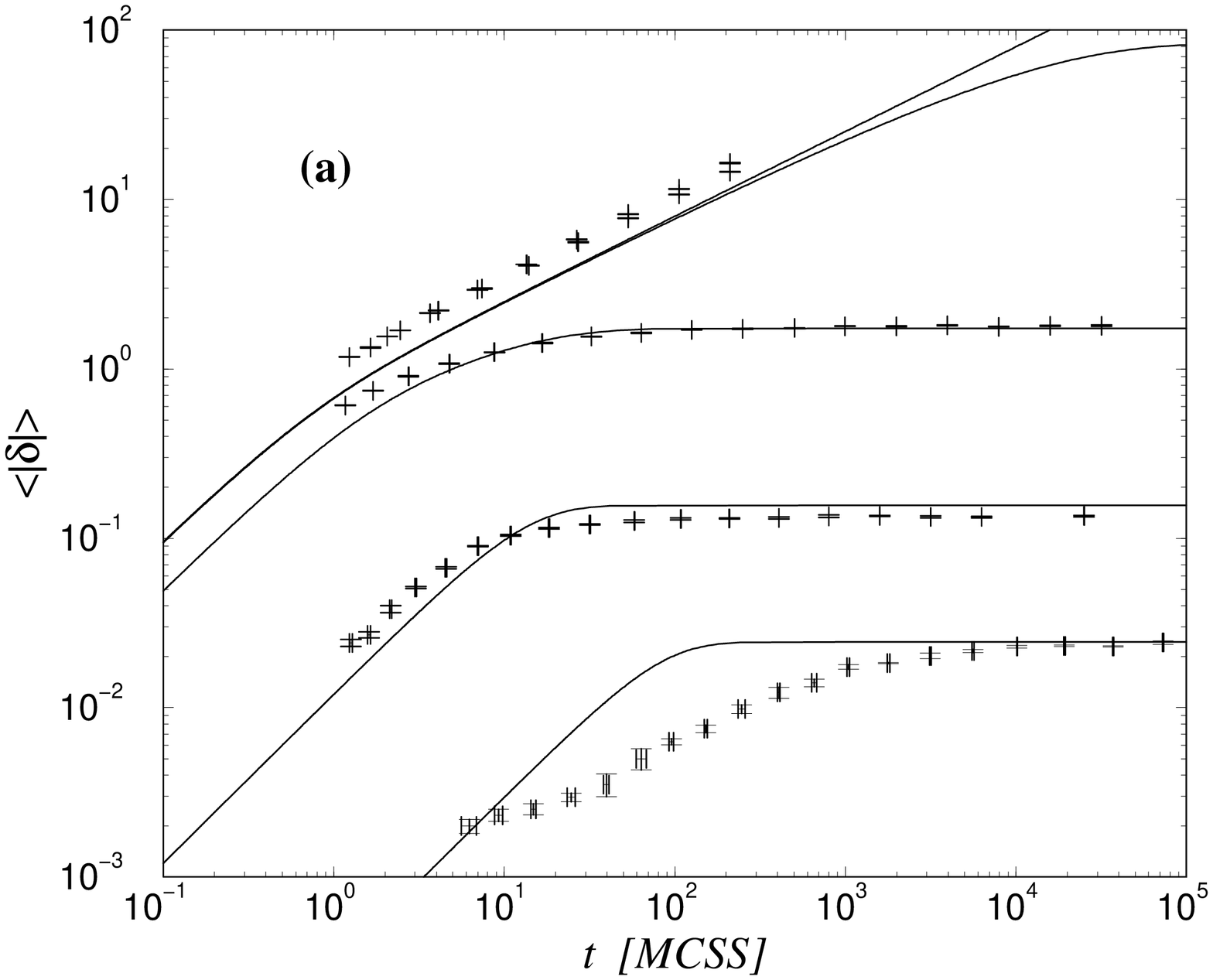} \\
\includegraphics[angle=270,width=.48\textwidth]{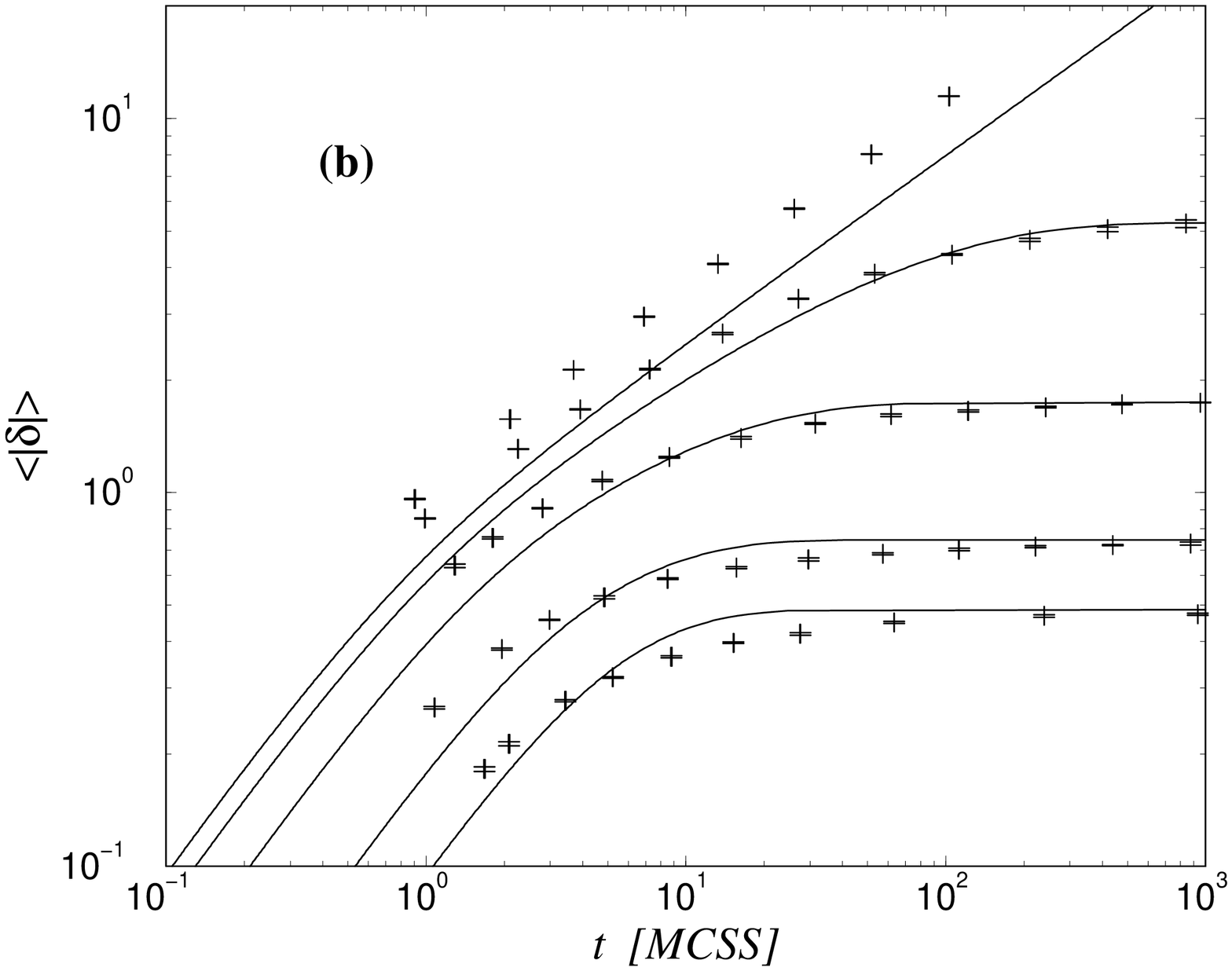}
\caption[]{
The mean step height $\langle | \delta | \rangle$, shown vs time 
$t$ (in MCSS) as predicted by the mean-field equation of motion for the 
single-step pdf, Eq.~(\protect\ref{eq:eom}) 
(solid curves), and by dynamic MC 
simulations (crossed error bars indicating statistical standard errors in 
$t$ and $\langle | \delta | \rangle$). 
From bottom to top, the results shown are 
for $H/J = 0$, 1, 2, 3, and 10 [MC results for $H/J = 10$ in part (b) only]. 
(a)
$T = 0.2T_c$. 
(b)
$T = 0.6T_c$. 
}
\label{fig:tdep}
\end{figure}

For comparison with the MC simulations, 
we solved the mean-field equation of motion for the single-step pdf, 
Eq.~(\ref{eq:eom}), numerically 
by a first-order iterative scheme with a time step of 10$^{-4}$~MCSS 
(shorter time steps made no discernible difference). 
Both the simulations and the solution of the equation of motion 
were started from a sharp interface at $t=0$. 
Results for fields between 0 and 10$J$ are shown in Fig.~\ref{fig:tdep}.

In general, we find overall qualitative agreement between the 
simulations and the equation of motion. For $T=0.2T_c$ and $|H| \leq 2J$, 
both methods have reached a common stationary value by $t = 10 \,000$~MCSS, 
while for $T=0.6T_c$ and $|H| \leq 3J$, stationarity is reached by 
$t = 1 \,000$~MCSS. Our choice of 5\,000~UPS as ``warm-up time'' 
in our studies of the stationary properties, corresponding to 
at least $10 \,000$~MCSS, is thus well justified. 
However, there are significant quantitative differences between the simulation 
results and the solution of Eq.~(\ref{eq:eom}) for early times. We believe this 
indicates that the mean-field assumption of statistically independent step 
heights is not well justified until the interface structure has been 
``randomized'' through a sufficient number of updates. For the extremely 
strong field, $H = 10J$, 
${\cal W}(\delta \rightarrow \delta \pm 1) \approx 
{\cal W}(\delta \pm 1 \rightarrow \delta) \approx 1/2$, so that the evolution 
of the interface width is essentially diffusional and 
$\langle | \delta | \rangle \propto t^{1/2}$. Both methods agree with this 
result, although the amplitude for the MC simulation is larger than 
predicted by the mean-field equation of motion. 

\subsection{Single-step probability densities}
\label{sec:MCss}

\begin{figure}[ht] 
\includegraphics[angle=270,width=.48\textwidth]{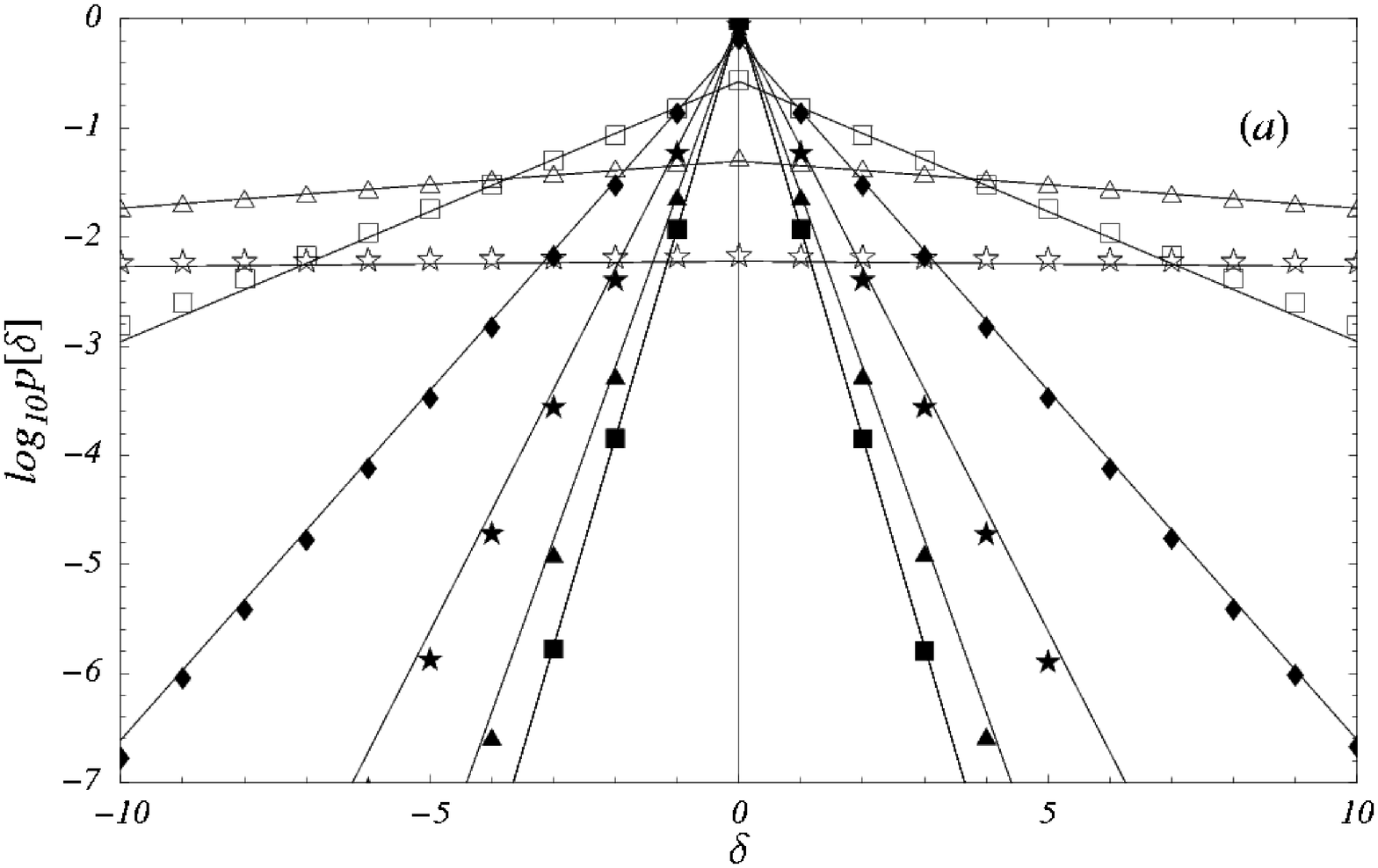} \\
\includegraphics[angle=270,width=.48\textwidth]{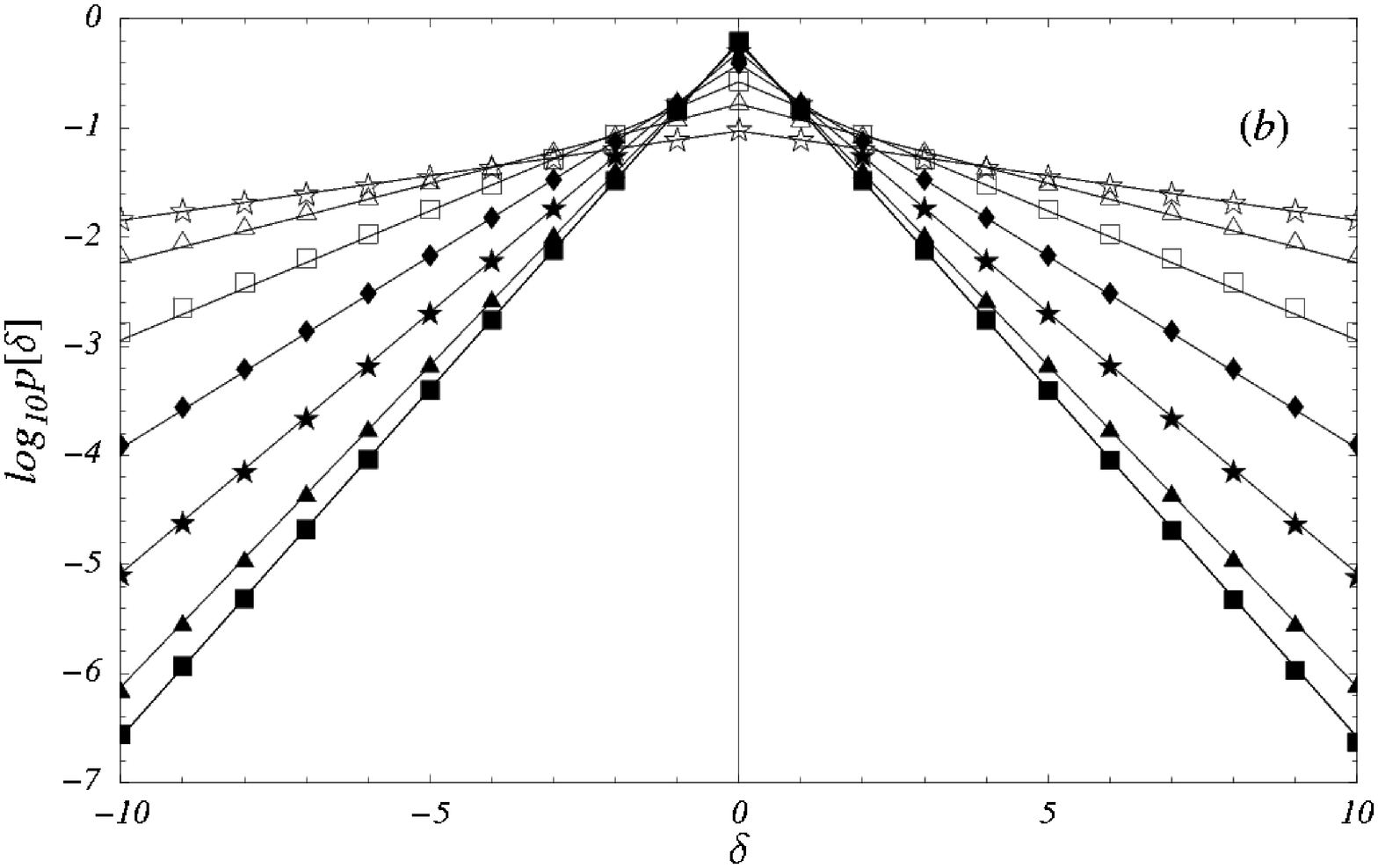}
\caption[]{
MC (data points) and analytical (solid lines) results for the stationary 
single-step pdf, shown on logarithmic scale vs $\delta$. The fields are 
$H/J = 0$ (filled squares), 0.5 (filled triangles), 1.0 (filled stars), 
1.5 (filled diamonds), 2.0 (empty squares), 2.5 (empty triangles), 
and 3.0 (empty stars). 
(a) $T=0.2T_c$. 
(b) $T=0.6T_c$. 
}
\label{fig:ss}
\end{figure}

Single-step pdfs were obtained by MC simulation at $T=0.2T_c$ and 
$0.6T_c$ for $\phi = 0$ and 
several values of $H$ between 0 and 3.0$J$. The simulation 
results for $p[\delta]$ are shown in Fig.~\ref{fig:ss}, together with 
the theoretical result, 
Eq.~(\ref{eq:step_pdf}) with $X(T,H)$ from Eq.~(\ref{eq:XG}). 
For both temperatures, the agreement 
is excellent in the whole range of $\delta$ and $H$ shown. 

\begin{figure}[ht] 
\includegraphics[width=.48\textwidth]{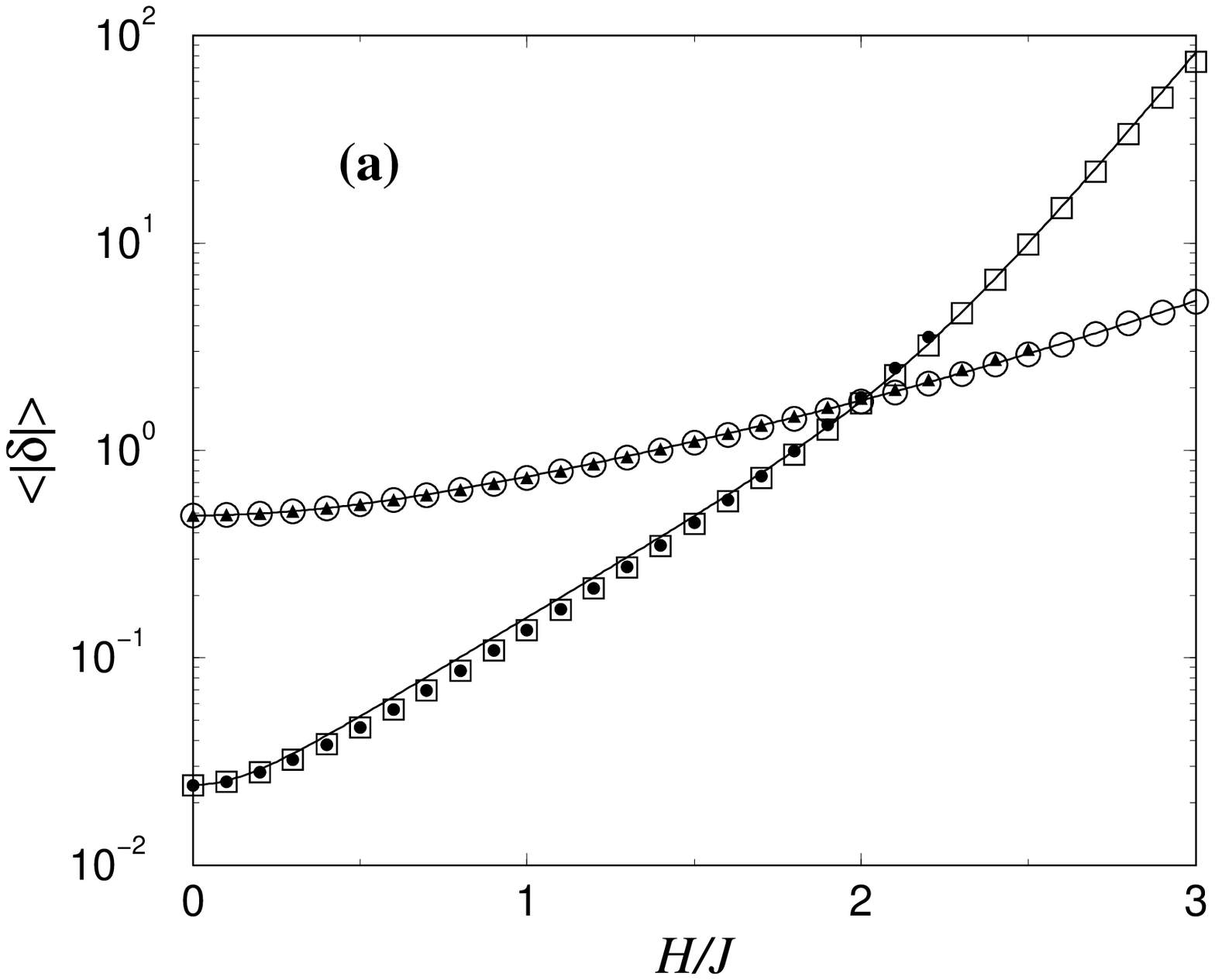} \\
\includegraphics[width=.48\textwidth]{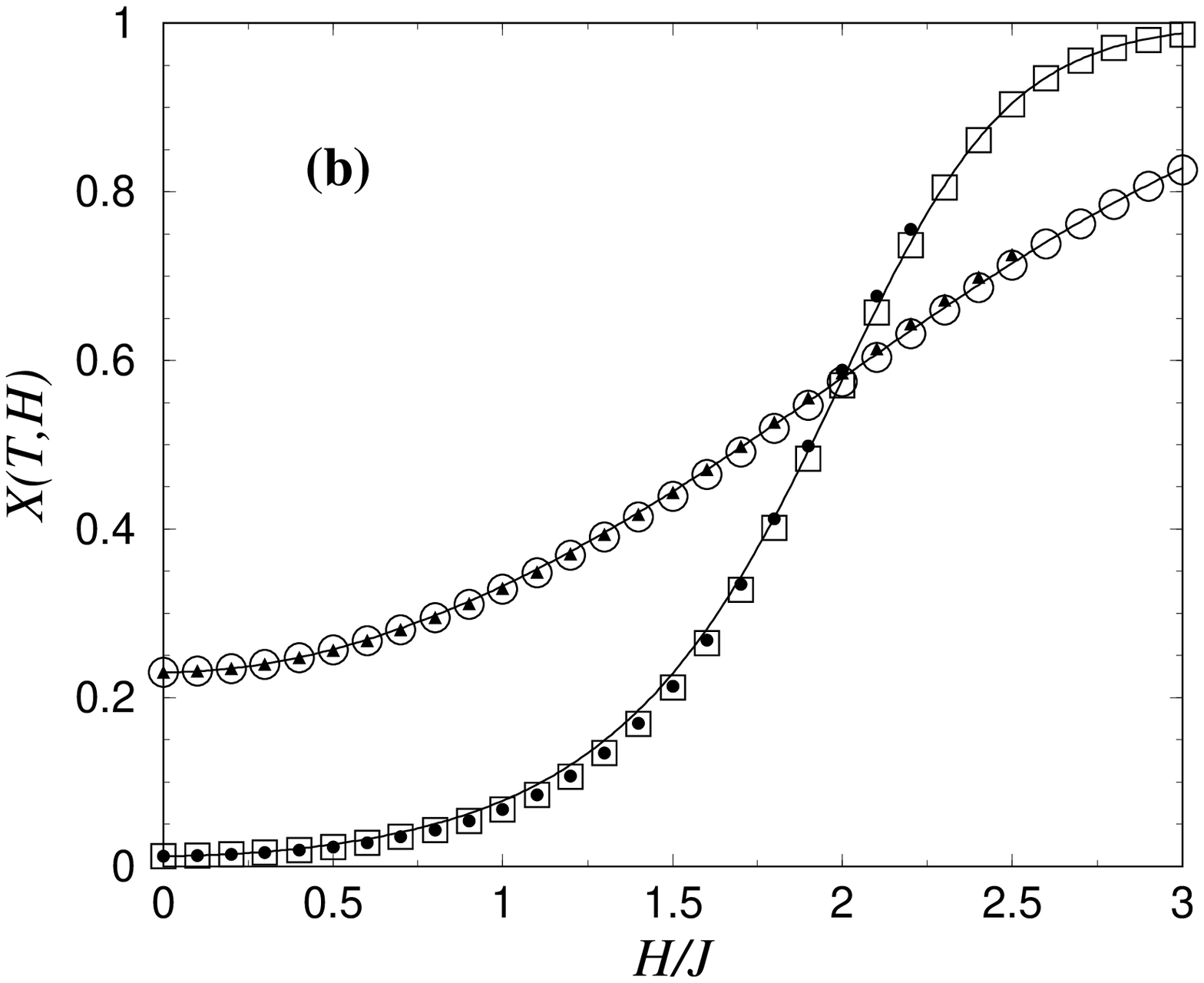}
\caption[]{
(a)
Average stationary step height 
$\langle | \delta | \rangle$ vs $H$ for $\phi$=0 at $T$=$0.2T_c$ and~0.6$T_c$.
The curves represent the theoretical result. 
The MC data were obtained directly by summation 
over the simulated single-step pdfs (filled symbols) and from the probability 
of zero step height (empty symbols). 
See text for details. 
Curve with filled circles and empty squares: 0.2$T_c$. 
Curve with filled triangles and empty circles: 0.6$T_c$. 
(b) 
The stationary pdf width parameter $X(T,H)$ vs $H$: the analytical result 
Eq.~(\protect\ref{eq:XG}) (curves) and 
estimates based on MC simulation results with $\phi$=0 
for $p[0]$ (empty symbols) 
and for $\langle | \delta | \rangle$ (filled symbols). 
See text for details. 
Curves and symbols have the same interpretations as in (a). 
In this and all the following figures, the statistical uncertainty is 
much smaller than the symbol size. 
}
\label{fig:Xss}
\end{figure}

A simple comparison between the analytical and simulation results is given in 
Fig.~\ref{fig:Xss}(a), which shows $\langle | \delta | \rangle$ vs $H$ for 
$\phi = 0$ at $T = 0.2T_c$ and~0.6$T_c$. 
The solid curves represent the theoretical result obtained by summation
of Eq.~(\ref{eq:step_pdf}), 
$\langle | \delta | \rangle = 2X/\left( 1-X^2 \right)$, 
with $X$ from Eq.~(\ref{eq:XG}).  There is excellent 
agreement between the theoretical field dependence and the MC data. 
Additional confirmation of the form of 
the single-step pdf, Eq.~(\ref{eq:step_pdf}), is obtained from the simulation 
results by calculating  
$\langle | \delta | \rangle$, both directly by summation over the 
numerically obtained pdf and from the probability 
of zero step height as 
$\langle | \delta | \rangle = \left\{ p[0]^{-1} - p[0] \right\}/2$.  

A slightly 
different way to check the agreement between the analytical predictions and 
the simulation results for the single-step pdf, is to compare $X(T,H)$ as 
given by Eq.~(\ref{eq:XG}) with the same quantity obtained from the simulations
under the assumption that Eq.~(\ref{eq:step_pdf}) holds.
From this equation for the pdf, using $Z(0)$ from Eq.~(\ref{eq:Z0}), 
it follows that $X$ is given in terms of $p[0]$ as 
$X = \{1-p[0]\}/\{1+p[0]\}$, and 
in terms of $\langle | \delta | \rangle$ as 
$X = \sqrt{1 + \langle | \delta | \rangle^{-2}} - 
\langle | \delta | \rangle^{-1}$. Both these MC estimates for $X(T,H)$ 
are shown in Fig.~\ref{fig:Xss}(b). Again, the agreement is excellent. 
The slight deviations of the estimate based on $\langle | \delta | \rangle$ for 
large $H$ are probably due to the fact that data were only recorded in 
separate bins for $| \delta | < 64$, so that the calculated 
average becomes inaccurate whenever higher steps cannot be ignored. The 
estimate based on $p[0]$ does not suffer from this problem. However, 
the slight discrepancy between both MC estimates on the one hand and 
Eq.~(\ref{eq:XG}) on the other, which is seen between $H/J = 0.5$ and 2 for 
$T = 0.2T_c$, is probably a real effect.

\subsection{Interface velocities}
\label{sec:MCi}

\begin{figure}[ht] 
\includegraphics[angle=270,width=.48\textwidth]{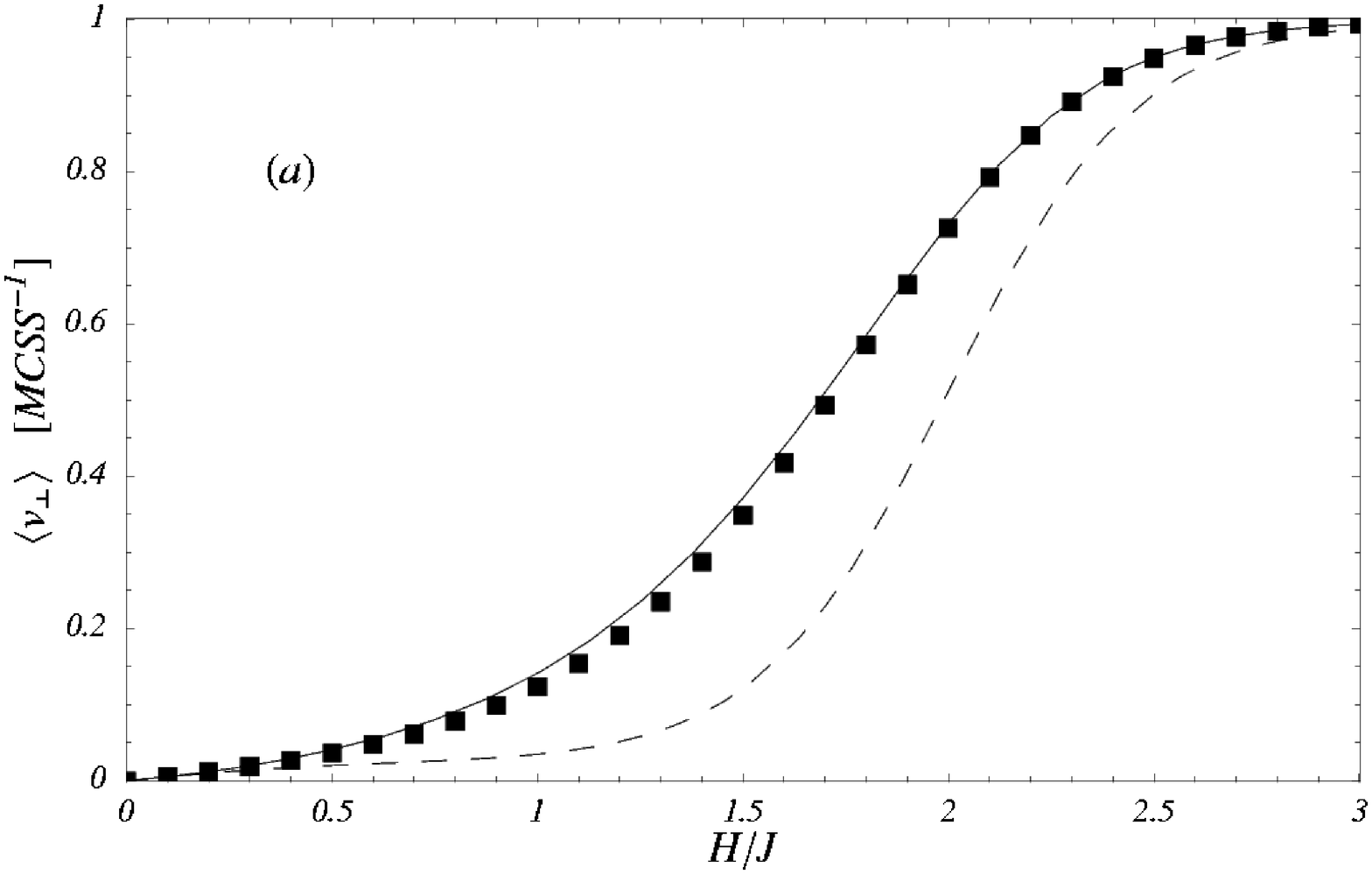} \\
\includegraphics[angle=270,width=.48\textwidth]{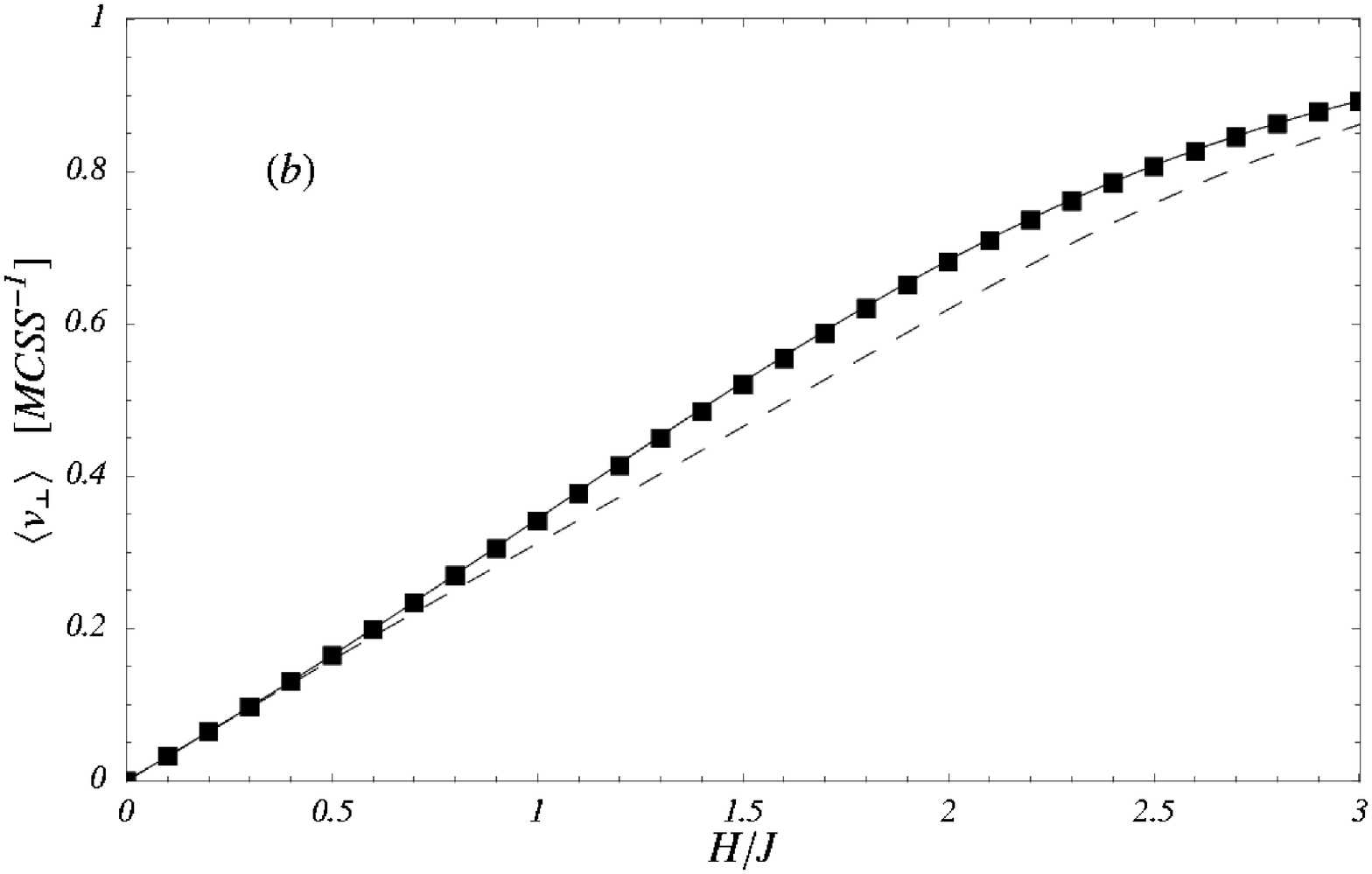}
\caption[]{
The average stationary 
normal interface velocity $\langle v_\perp \rangle$ vs $H$ for 
$\phi = 0$. The MC results are shown as data points, the linear-response results 
as dashed curves, and the nonlinear-response results as solid curves. 
(a) 
$T= 0.2T_c$. 
(b) 
$T= 0.6T_c$. 
}
\label{fig:vvsH}
\end{figure}

In this section we compare the simulated interface velocities 
with the analytical approximation, Eq.~(\ref{eq:totalv}). 
Figure~\ref{fig:vvsH} shows the normal velocity vs $H$ for 
$\phi = 0$. Included are both the linear-response approximation (i.e., 
$X= e^{-2 \beta J_x}$) and the nonlinear-response result with $X(T,H)$ from
Eq.~(\ref{eq:XG}). Overall, there is excellent agreement between the 
MC results and the nonlinear-response theory, while the linear-response
approximation seriously underestimates the velocity, especially at the lower 
temperature. As for the Ising model with hard Glauber dynamics studied in 
Ref.~\cite{RIKV00}, these results show that the latter approximation is clearly 
inadequate, and we include no further linear-response results in this 
paper. A slight disagreement between the simulations and the analytical 
predictions is seen at 0.2$T_c$ in the same range of field 
values as for $X(T,H)$ obtained from the single-step pdf in Sec.~\ref{sec:MCss}. 

\begin{figure}[ht] 
\includegraphics[width=.38\textwidth,angle=270]{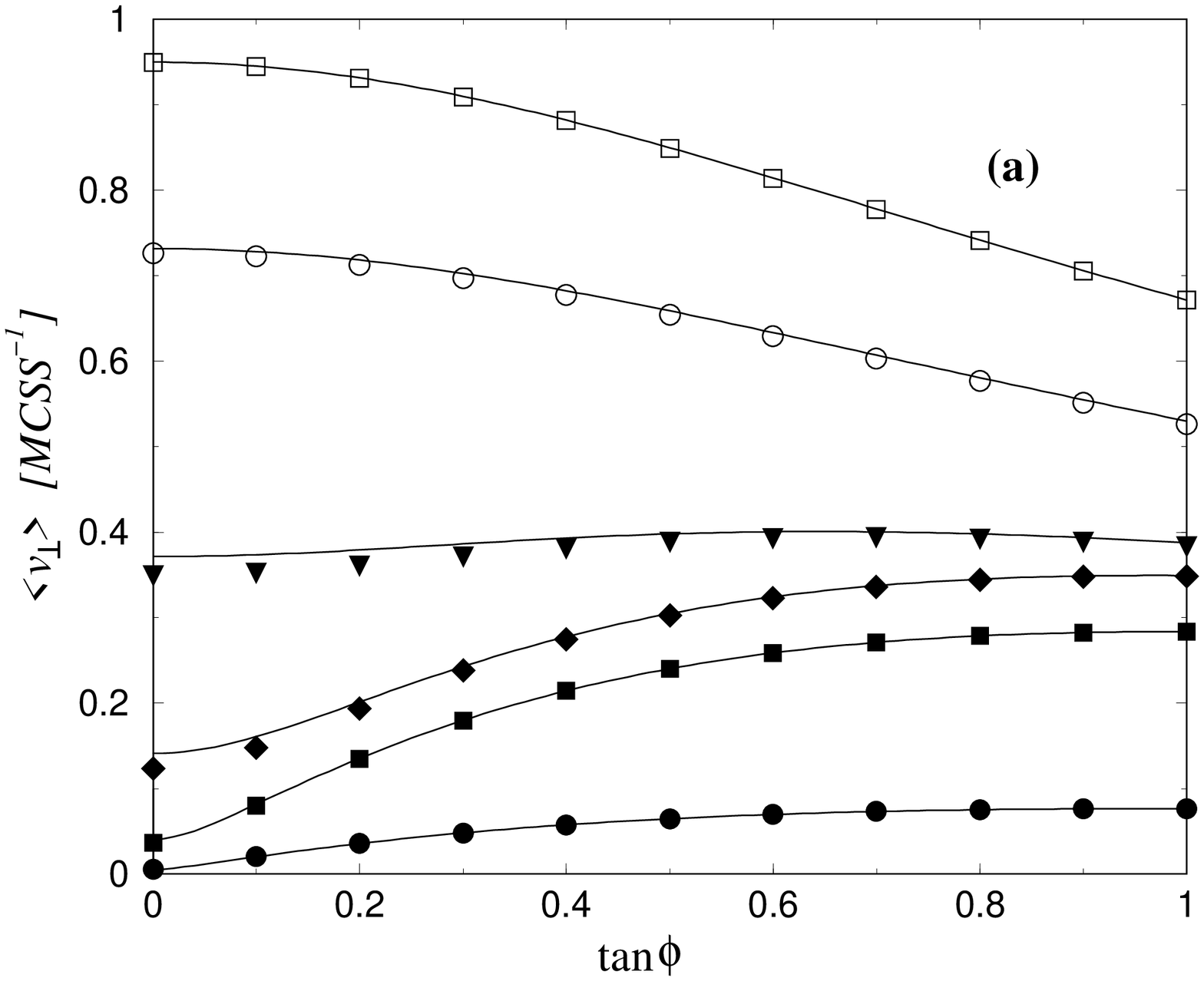} \\
\includegraphics[width=.38\textwidth,angle=270]{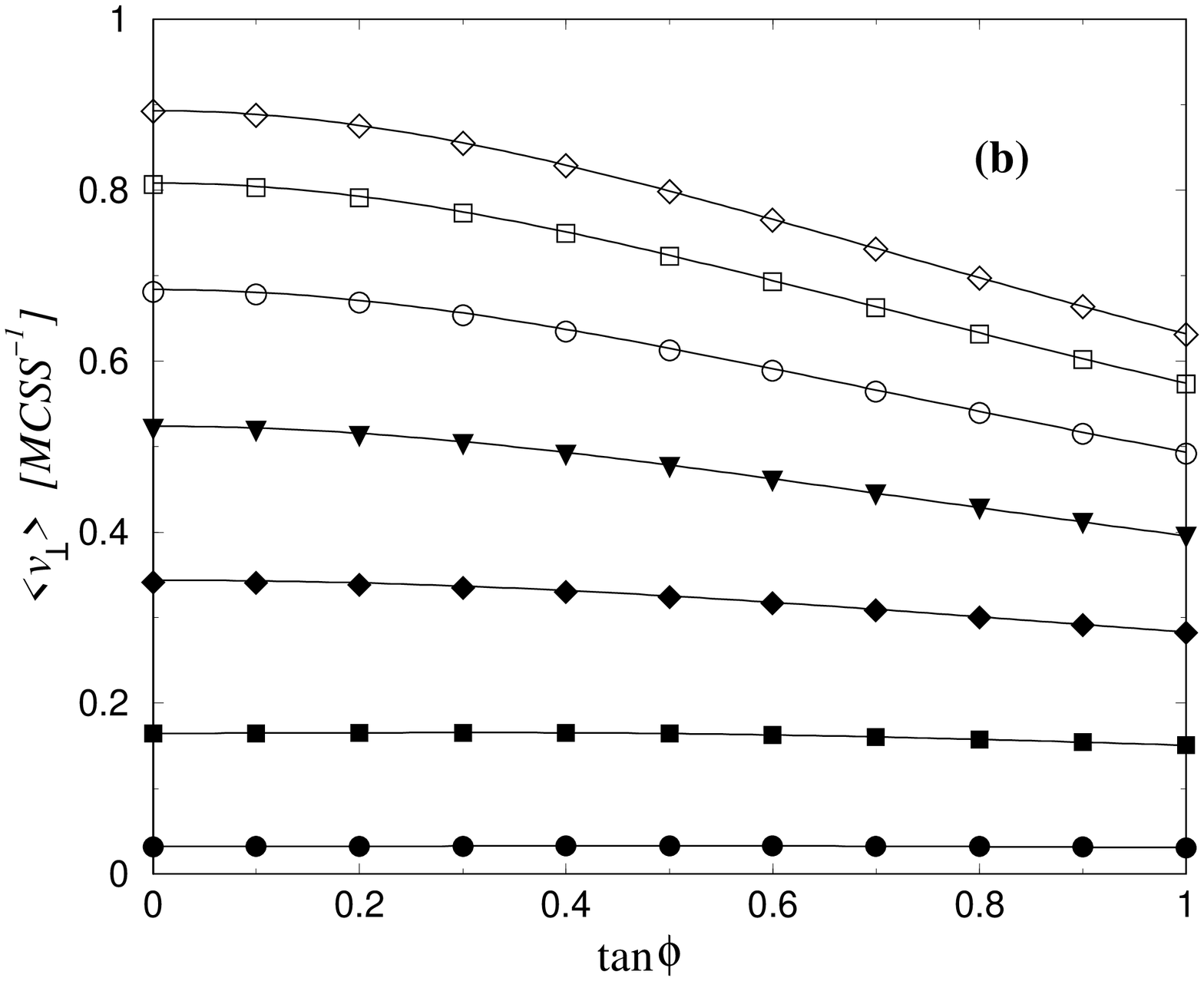} \\
\includegraphics[width=.43\textwidth,angle=0]{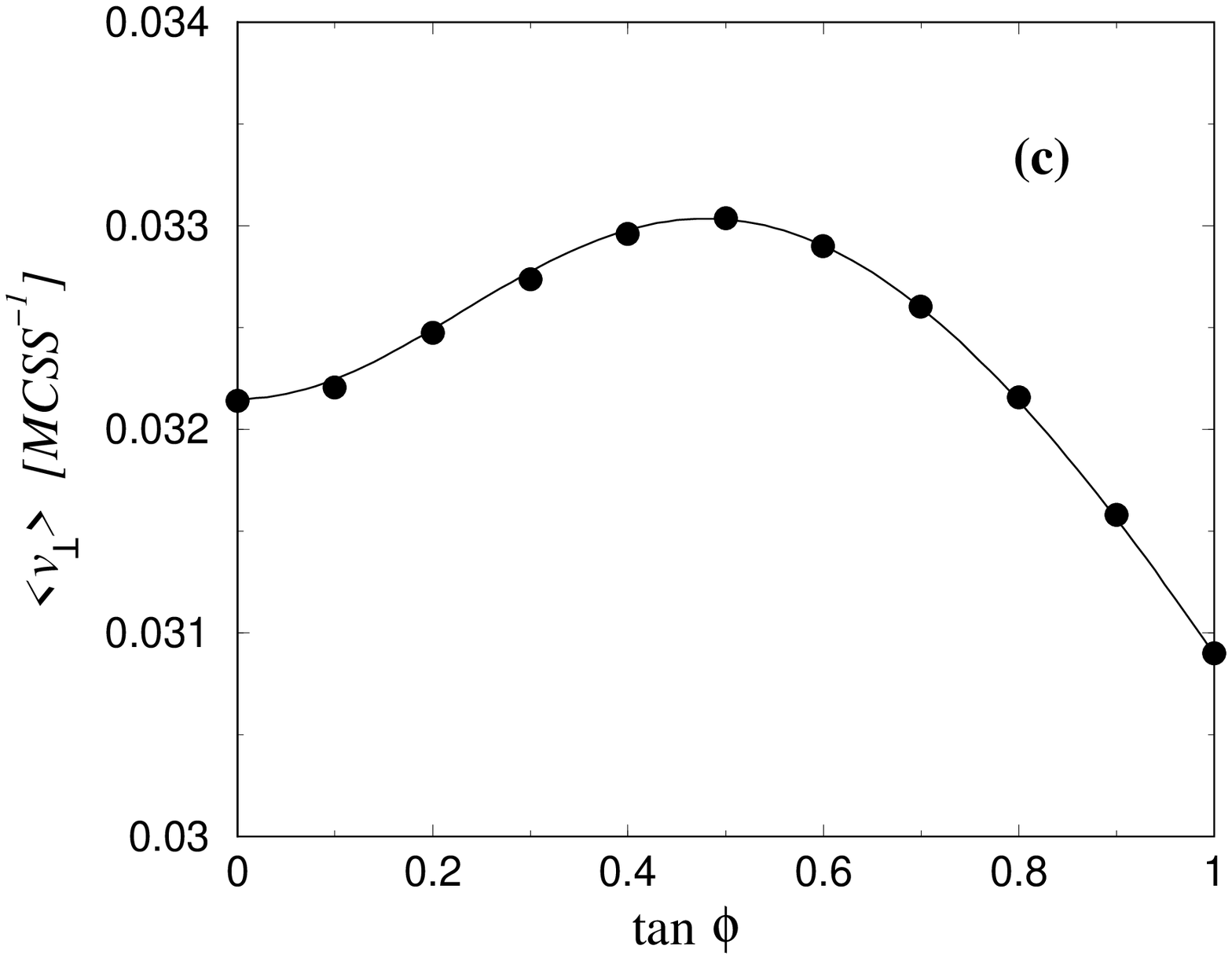}
\caption[]{
The average stationary 
normal interface velocity $\langle v_\perp \rangle$ vs $\tan \phi$ 
for several values of $H$. 
MC data are represented by data points, and analytical results by solid curves. 
(a) 
$T= 0.2T_c$. 
(b) 
$T= 0.6T_c$. 
In parts (a) and (b), the values of $H/J$ are (from below to above): 
0.1, 0.5, 1.0, 1.5, 2.0, 2.5, and 3.0 [in (b) only]. 
(c)
$H/J=0.1$ for $T= 0.6T_c$, shown on a magnified scale to reveal the nonmonotonic
dependence on $\tan \phi$ for this very weak field. 
}
\label{fig:vvsA}
\end{figure}

The dependence of the normal velocity on the tilt angle $\phi$ is shown in 
Fig.~\ref{fig:vvsA} for several values of $H/J$ between 0.1 and 3.0. 
At $T=0.2T_c$ the anisotropy undergoes a gradual change from increasing with 
$\phi$ in agreement with the polynuclear growth model at 
small angles and the single-step model for larger angles at weak fields, to 
Eden-type inverse anisotropy \cite{DHAR86,MEAK86B,HIRS86} at strong fields 
[Fig.~\ref{fig:vvsA}(a)]. At $T=0.6T_c$, on the other hand, 
inverse anisotropy is found for 
the stronger fields, growing gradually more pronounced with increasing $H$ 
[Fig.~\ref{fig:vvsA}(b)], 
while for the weakest fields studied 
the velocity is nonmonotonic in $\tan \phi$ [Fig.~\ref{fig:vvsA}(c)]. 
The agreement between the simulations and the analytical results is excellent 
everywhere. 

\begin{figure}[ht] 
\includegraphics[angle=270,width=.48\textwidth]{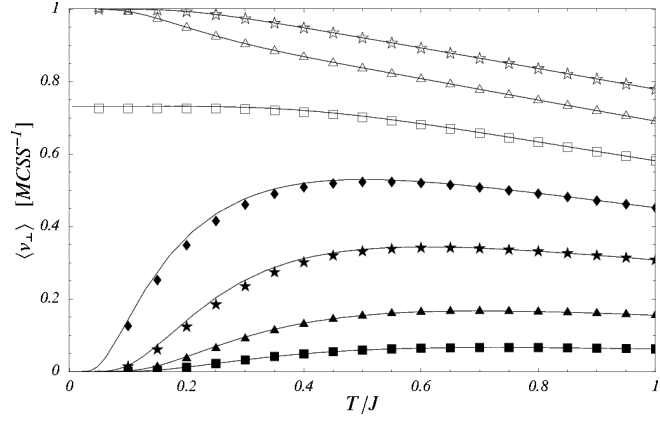}
\caption[]{
The average stationary 
normal interface velocity $\langle v_\perp \rangle$ vs $T$ 
for $\phi = 0$. MC data are represented by data points, 
and analytical results by solid curves. 
From below to above, the values of $H/J$ are 
0.2, 0.5, 1.0, 1.5, 2.0, 2.5, and 3.0. 
}
\label{fig:vvsT}
\end{figure}

The temperature dependence of the normal interface velocity is shown in 
Fig.~\ref{fig:vvsT} for several values of $H/J$ between 0.2 and 3.0. 
The agreement between the simulations and the analytical results is excellent 
everywhere. 

\subsection{Spin-class populations and skewness}
\label{sec:MCc}

A closer look at the performance of the mean-field approximation for the 
interface structure  
is provided by the mean spin-class populations. The analytical predictions 
for the class populations are 
based on the assumption that different steps are statistically independent. 
A comparison of the simulation results with the analytical 
predictions therefore gives a way of testing this assumption. 

\begin{figure}[ht] 
\includegraphics[angle=270,width=.48\textwidth]{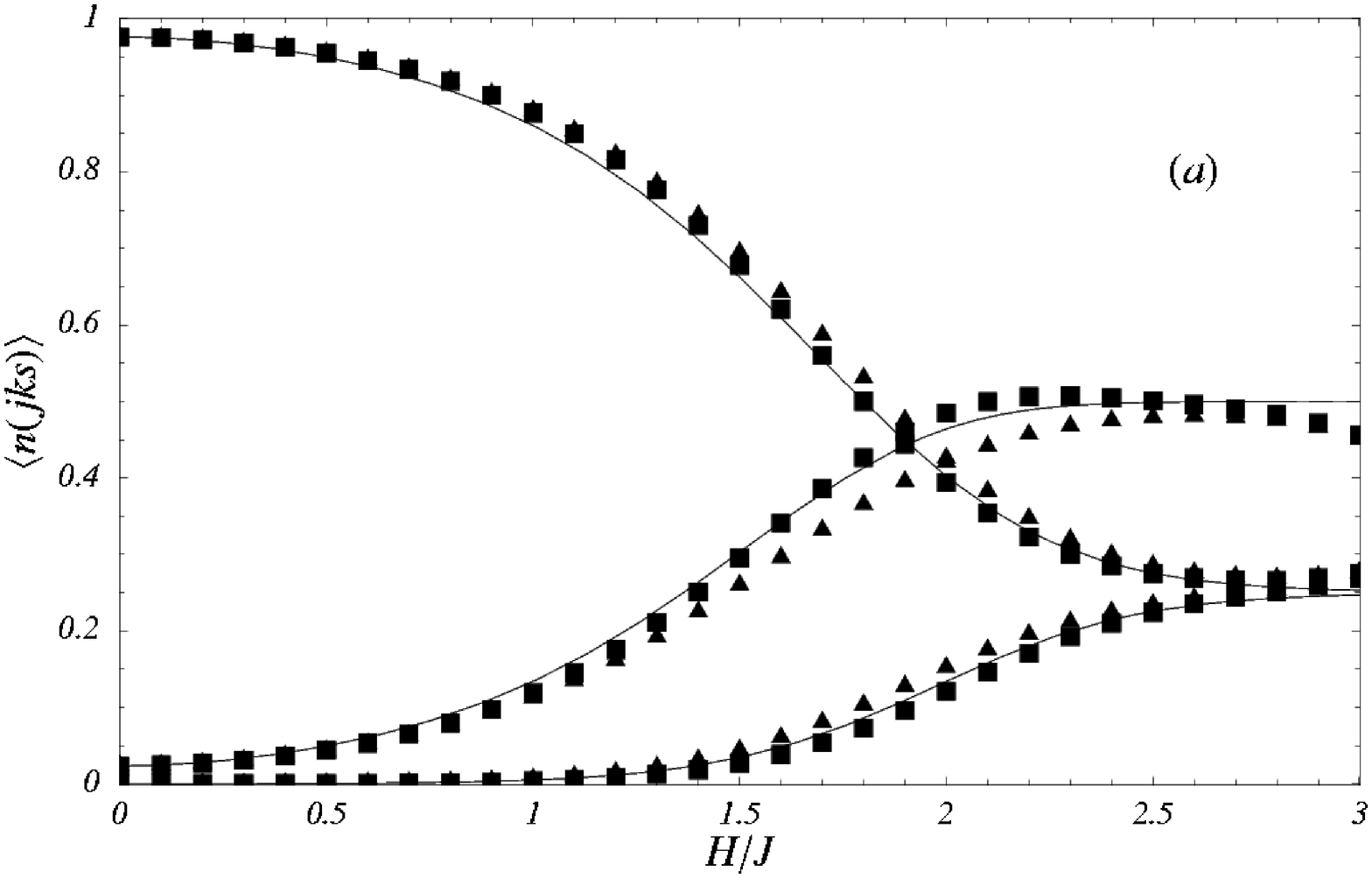} \\
\includegraphics[angle=270,width=.48\textwidth]{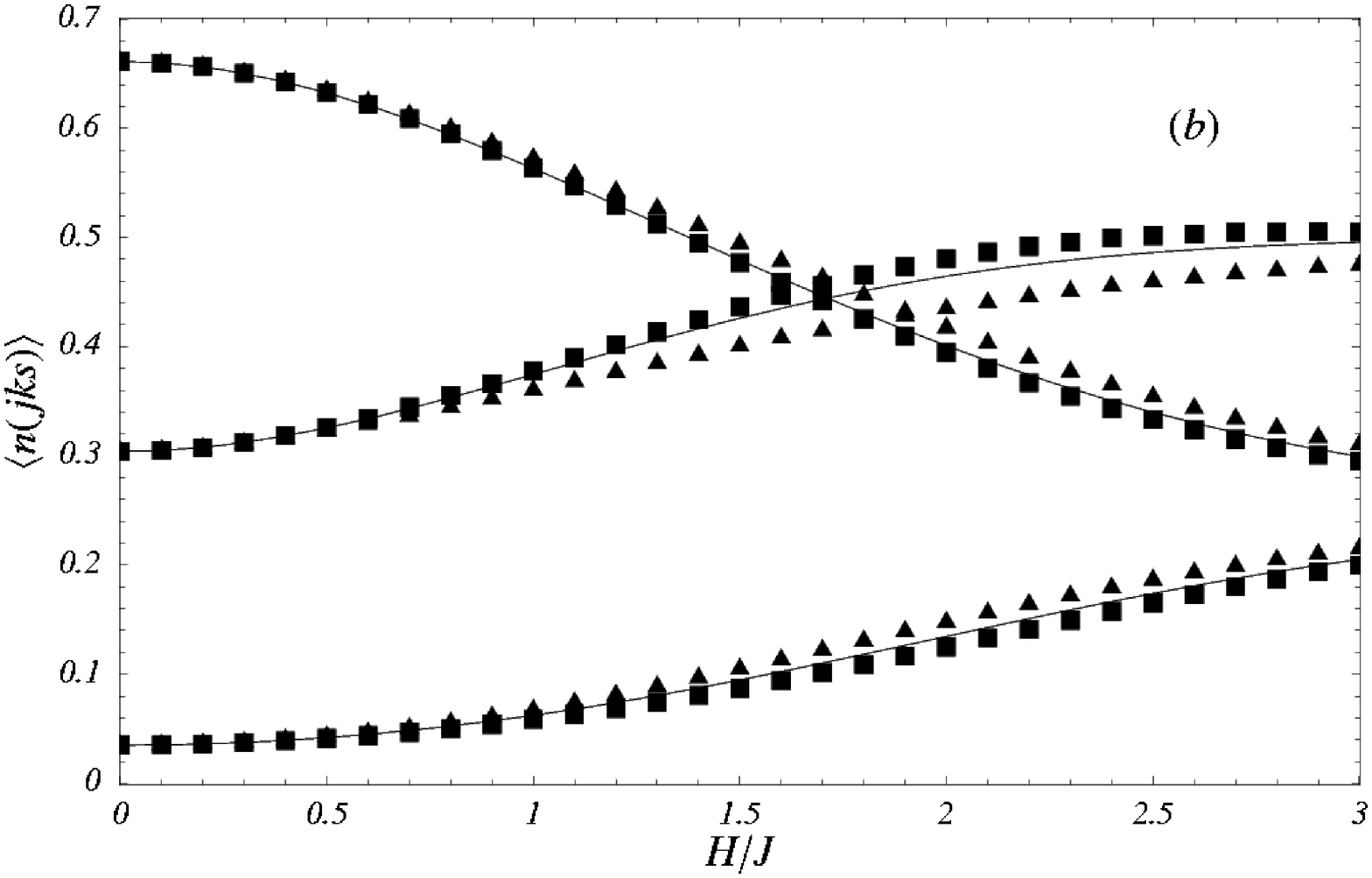}
\caption[]{
Mean stationary class populations $\langle n(jks) \rangle$ vs $H$ 
for $\phi = 0$. (a) $T=0.2T_c$. (b) $T=0.6T_c$. 
From top to bottom at the left edge of both parts, the classes are 
$01s$, $11s$, and $21s$ with squares representing MC data 
for $s = +1$ and triangles for $s=-1$. The analytic approximations are 
indicated by the solid curves. 
Note the different vertical scales in the two parts. 
}
\label{fig:cpop}
\end{figure}

The six mean class populations, $\langle n(01s) \rangle$, 
$\langle n(11s) \rangle$, 
and $\langle n(21s) \rangle$ with $s = \pm 1$ are shown vs $H$ 
in Fig.~\ref{fig:cpop} for $\phi = 0$ and $T=0.2T_c$ and $0.6T_c$. At both 
temperatures the 
analytical approximations follow the average of the populations 
for $s=+1$ and $s=-1$ quite well, but at intermediate fields in particular, 
the populations in front of the surface ($s = -1$) and behind it ($s=+1$) are 
distinctly different. 

\begin{figure}[ht] 
\includegraphics[angle=270,width=.48\textwidth]{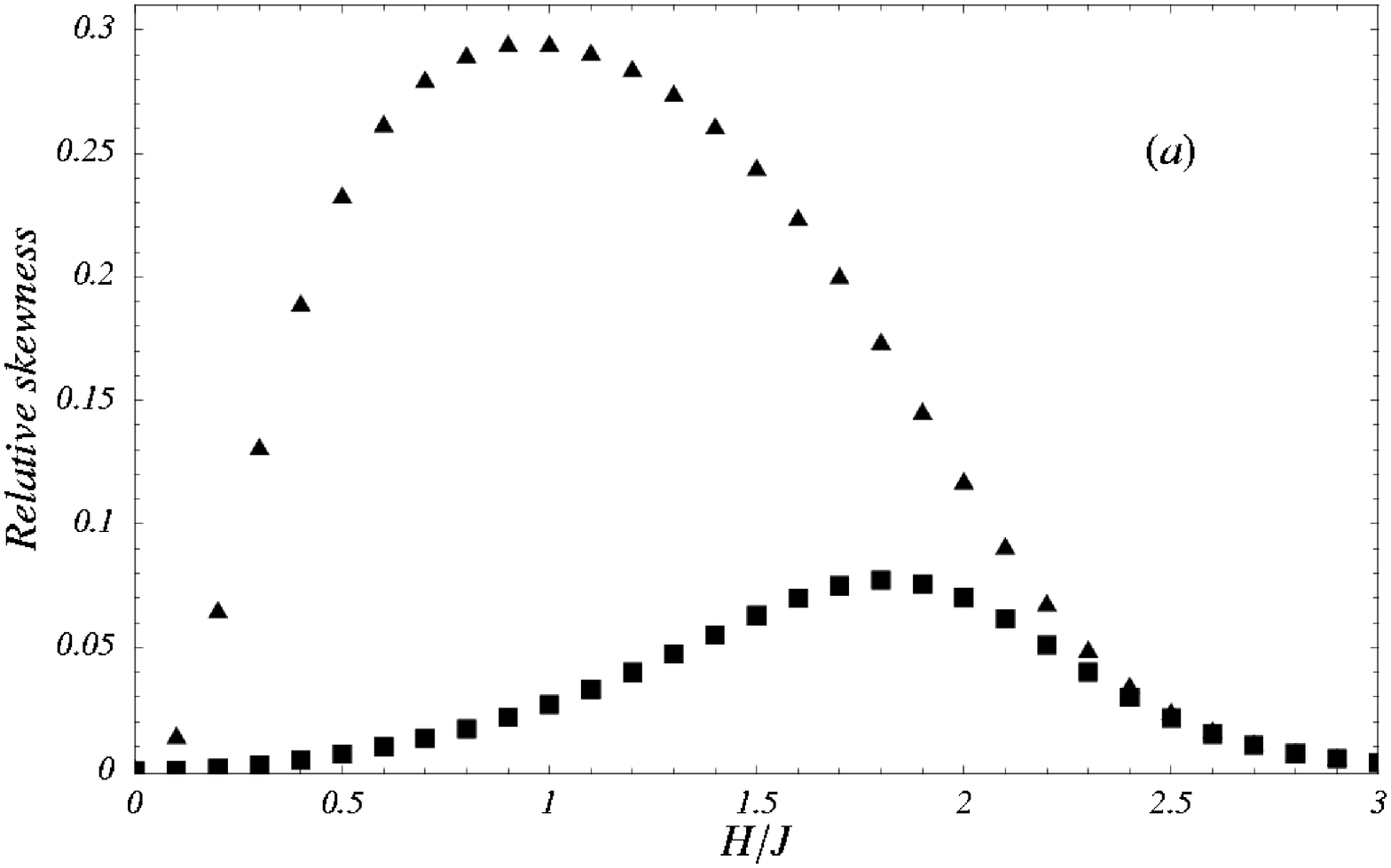} \\
\includegraphics[angle=270,width=.48\textwidth]{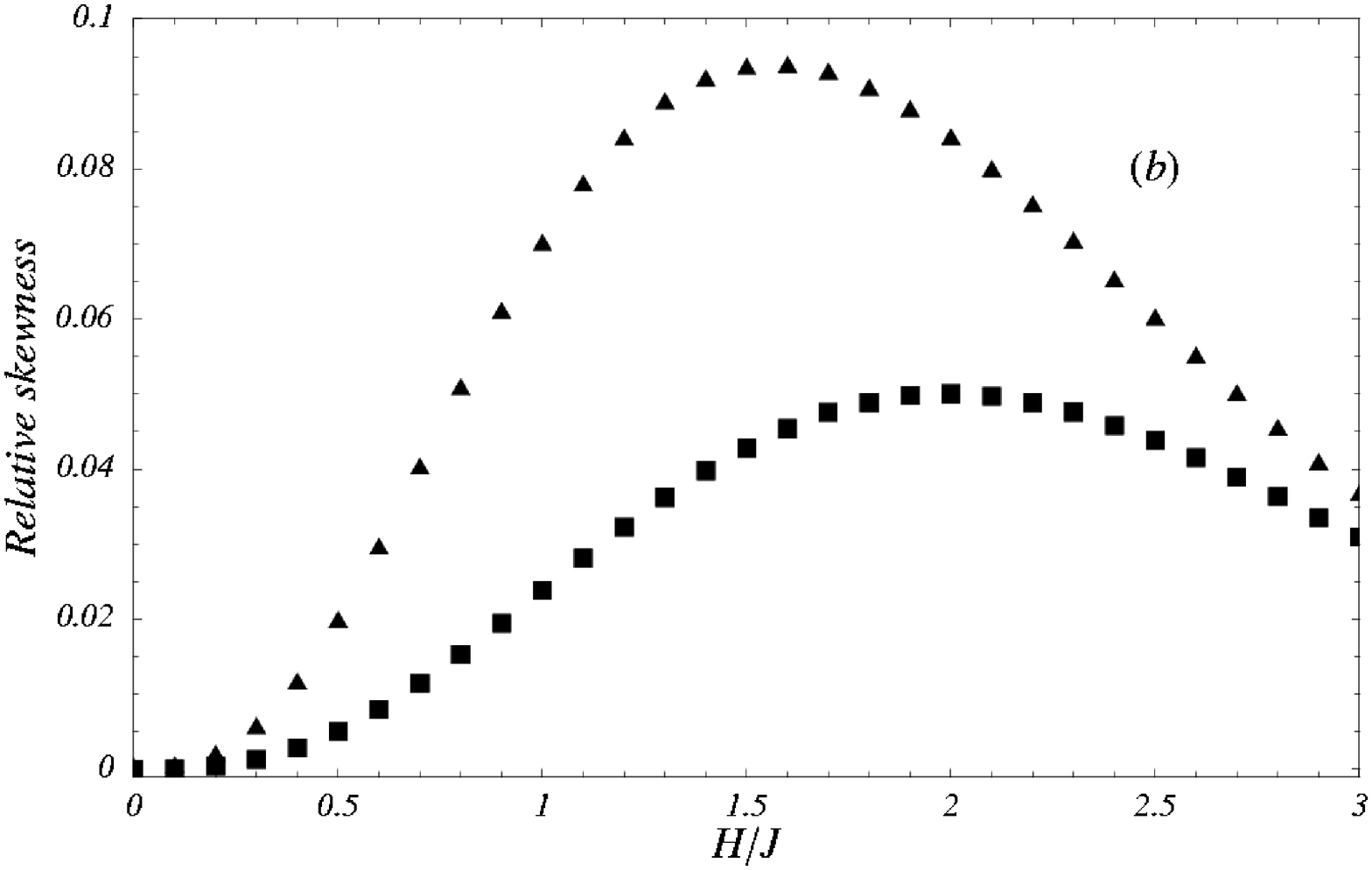}
\caption[]{
The two relative skewness parameters $\rho$ (triangles) and 
$\epsilon$ (squares), defined in 
Eqs.~(\protect\ref{eq:rho}) and~(\protect\ref{eq:epsi}), respectively. 
The parameters are shown vs $H$ 
for $\phi = 0$. (a) $T=0.2T_c$. (b) $T=0.6T_c$. 
Note the different vertical scales in the two parts. 
}
\label{fig:skew}
\end{figure}

The skewness between the spin populations on the leading and trailing edges 
of the interface are a consequence of short-range correlations between 
neighboring steps, and it is quite commonly observed in driven interfaces. 
This is the case, 
even when the {\it long-range\/} correlations vanish as they do for 
interfaces in the 
KPZ dynamic universality class, to which the present model belongs. 
Such skewness was also 
observed in our study of the Ising model in Ref.~\cite{RIKV00}, 
but in that case it was difficult to 
separate it from the effects of bubbles and overhangs. Skewness 
has also been 
observed in other SOS-type models, such as the body-centered SOS model 
studied by Neergaard and den~Nijs \cite{NEER97} and a model for the local time 
horizon in parallel MC simulations studied by Korniss et al.\ 
\cite{KORN00}. (However, no skewness is observed for the soft Glauber 
dynamic, a result which may be general for soft dynamics.) 
The correlations associated with the skewness generally lead to a broadening 
of protrusions on the leading edge (``hilltops''), while 
those on the trailing edge (``valley bottoms'') are sharpened \cite{NEER97}, 
or the other way around \cite {KORN00}. In terms of spin-class populations, 
the former corresponds to $\langle n(21-) \rangle > \langle n(21+) \rangle$ 
and $\langle n(11+) \rangle > \langle n(11-) \rangle$. The relative skewness 
can therefore be quantified by the two functions, 
\begin{equation}
\rho = \frac{\langle n(21-) \rangle - \langle n(21+) \rangle}
{\langle n(21-) \rangle + \langle n(21+) \rangle}
\;,
\label{eq:rho}
\end{equation}
introduced in Ref.~\cite{NEER97}, and 
\begin{equation}
\epsilon = \frac{\langle n(11+) \rangle - \langle n(11-) \rangle}
{\langle n(11+) \rangle + \langle n(11-) \rangle}
\;.
\label{eq:epsi}
\end{equation}
These two skewness parameters are shown together in Fig.~\ref{fig:skew}. 
The relative skewness is seen to be considerably stronger at the lower 
temperature. This temperature dependence is especially pronounced for $\rho$. 

\begin{figure}[ht] 
\includegraphics[angle=270,width=.38\textwidth]{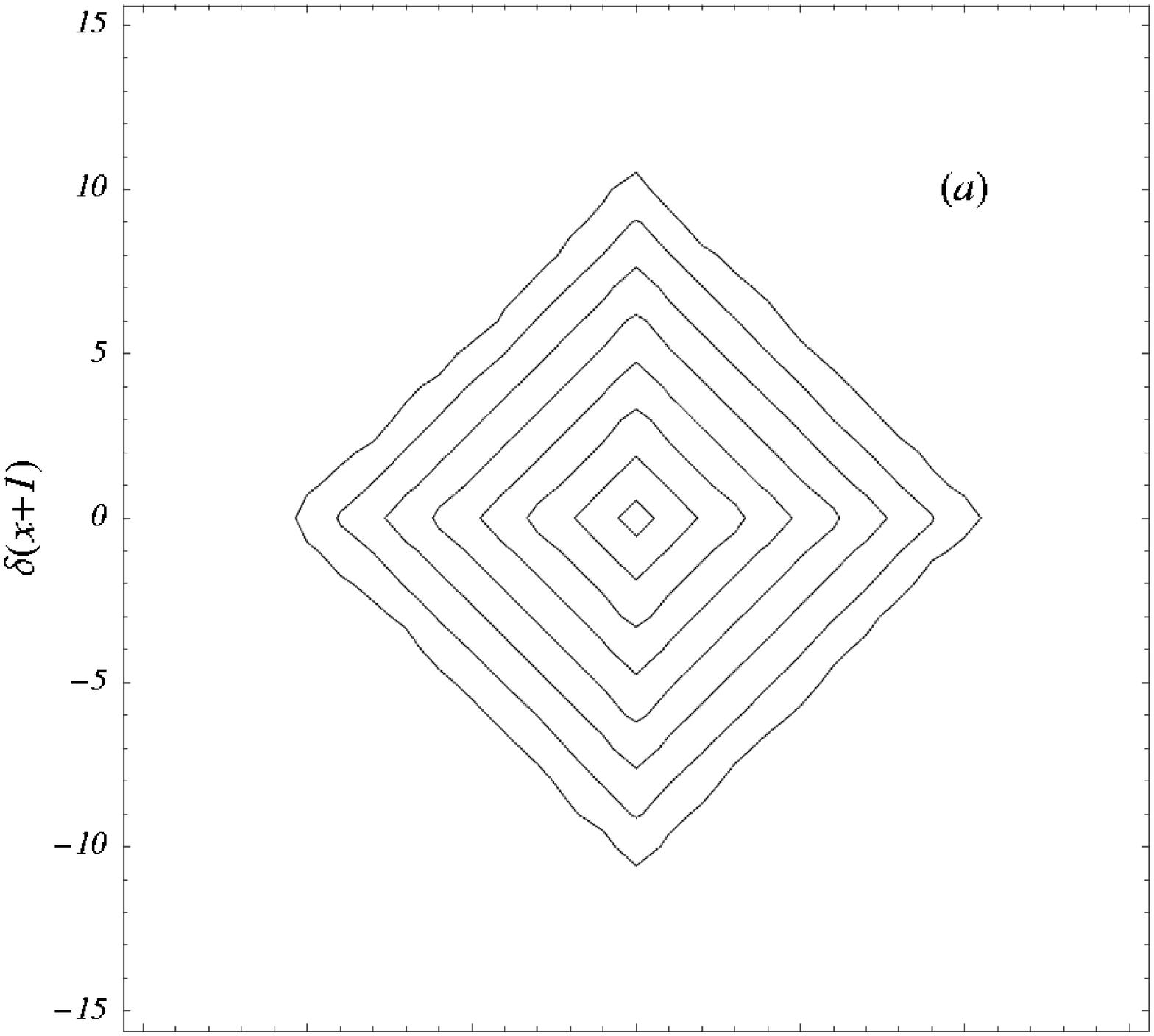} \\
\includegraphics[angle=270,width=.38\textwidth]{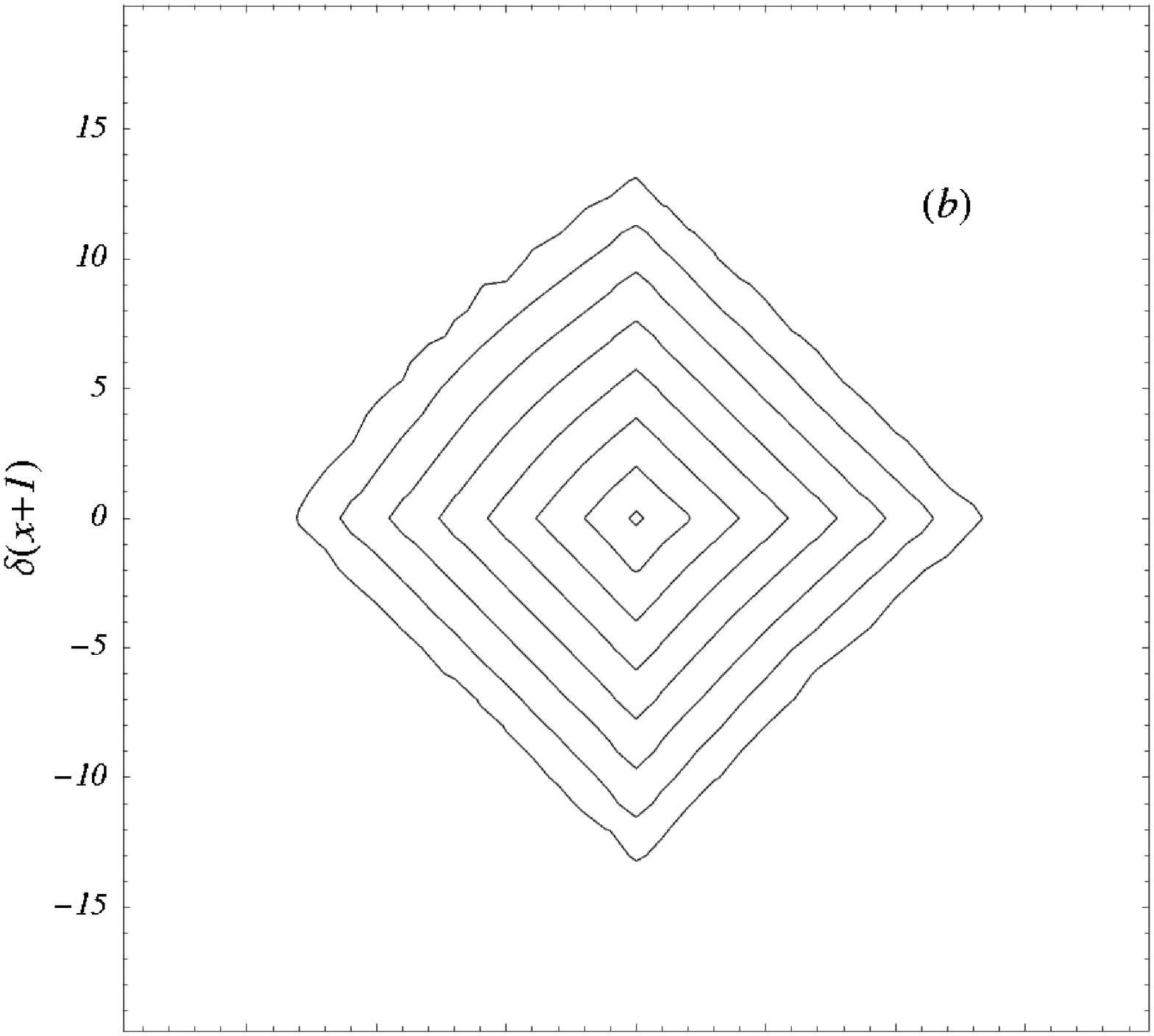} \\
\includegraphics[angle=270,width=.38\textwidth]{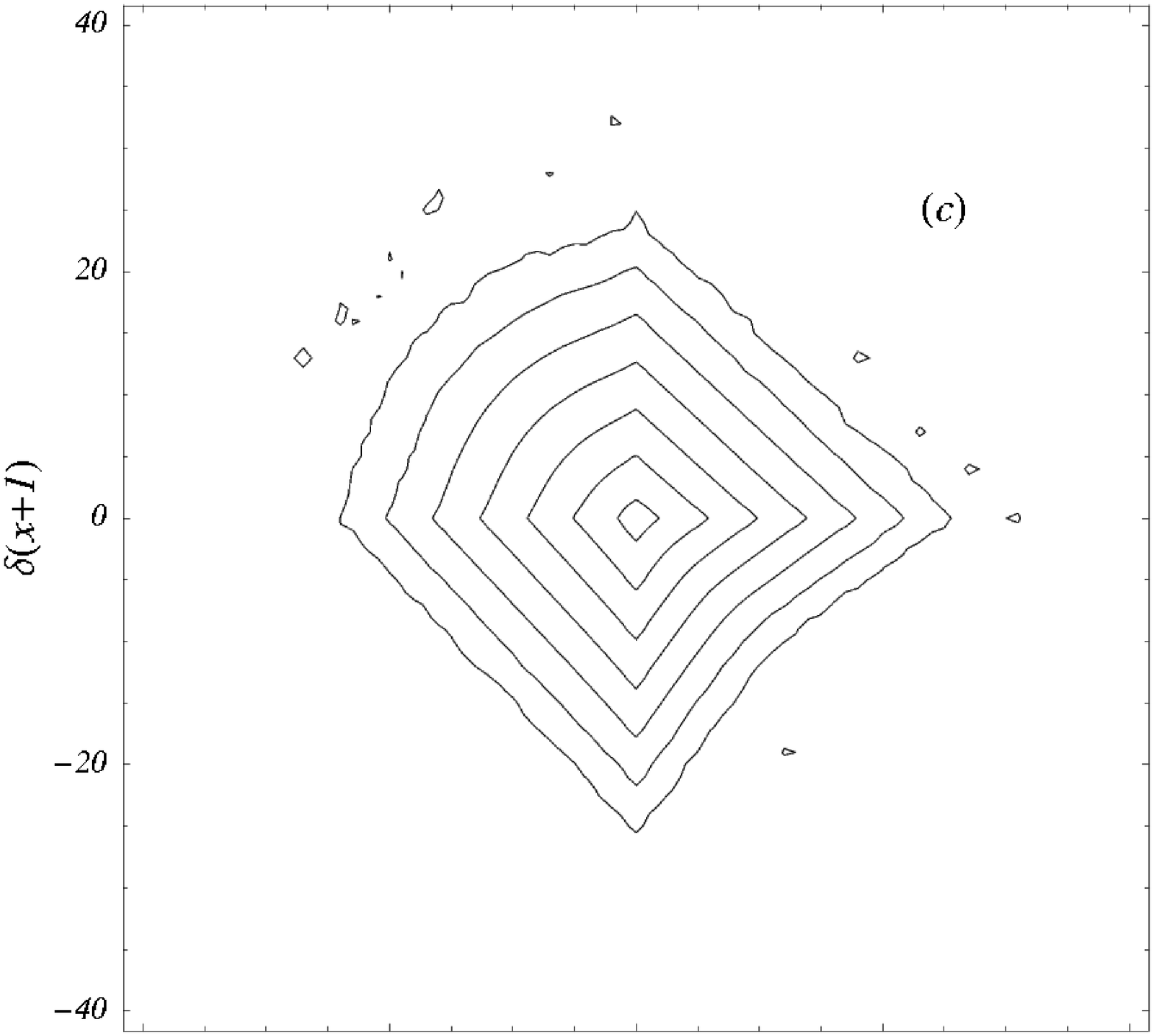}
\caption[]{
Contour plots of 
$\log_{10} p \left[ \delta(x) , \delta(x+1) \right]$ for $\phi=0$. 
(a)
$H/J = 0$. 
(b)
$H/J = 1.0$. 
(c)
$H/J = 2.0$.
Note the different scales in the three parts. 
See discussion in the text.  
}
\label{fig:contour}
\end{figure}

Yet another way to visualize the skewness is to consider the joint 
two-step pdf, $p \left[ \delta(x),\delta(x+1) \right]$. Logarithmic contour 
plots of this quantity for different values of $H$ 
are shown in Fig.~\ref{fig:contour} for $\phi=0$ at $T=0.6T_c$. It is clearly 
seen how the contours change with $H$. 
For $H$=0 a symmetric diamond shape with equidistant contours indicates 
statistical independence with single-step pdfs given by Eq.~(\ref{eq:step_pdf}).  
For stronger fields we find shapes that are elongated in the second 
quadrant [$\delta(x) < 0$, $\delta(x+1) > 0$] 
and foreshortened in the fourth quadrant [$\delta(x) > 0$, $\delta(x+1) < 0$]. 
This shape indicates that large negative $\delta(x)$ tend to be followed by 
large positive $\delta(x+1)$ (sharp valleys), 
while positive $\delta(x)$ tend to be followed by smaller negative 
$\delta(x+1)$ (rounded hilltops). 
We have not been successful in attempts to construct an analytical 
approximation which describes this evolution of the joint two-step pdf with 
increasing field. 

\section{Conclusion}
\label{sec:DISC}

In this paper we have considered in detail the 
microstructure of an unrestricted solid-on-solid (SOS) interface 
with Glauber dynamics, 
which is driven far from equilibrium by an applied field. 
The microstructure is of interest because it determines a number 
of interface properties, such as mobility and chemical reactivity. 
We adapted to this model a mean-field, 
nonlinear-response approximation previously developed for driven Ising 
interfaces without the SOS restriction \cite{RIKV00}.
In comparison to the Ising driven interface, which leaves bubbles of the unstable
phase in its wake and exhibits ``overhangs,'' the SOS interface is 
a relatively simple object. The absence of overhangs and of 
fluctuations in the stable and unstable phases (bubbles behind and in front of
the interface) makes the SOS interface more suitable for description
in terms of a mean-field type model. Moreover, unlike the Ising model,
in which there are several effects that simultaneously contribute to the
inaccuracy of the approximate treatment, the simpler SOS structure
makes it possible to identify the short-range correlations as the only
significant factor causing deviations between the true interface behavior 
and the mean-field theory.

To study the microstructure of the interface in detail, we investigated
the interface velocity as a function of driving field,
temperature, and angle relative to the lattice axes. We also
studied the local shape of the interface in terms of the spin-class
populations and the probability density for individual steps in the
interface. In essentially all cases we found excellent
agreement between our theoretical description of the stationary moving 
interface and the results of our dynamic MC simulations. 

The microstructure of the moving interface depends crucially on the details of 
the stochastic dynamics, and for the Glauber dynamic used here, 
the average height of a step in the interface was found to increase strongly 
with the applied field. Our theory 
predicts that this should be the case (with quantitative variations depending 
on the particular dynamic) for any dynamic in which the single-spin 
transition rates cannot be factorized 
into a part that depends only on interaction
energies and another that depends only on the applied field 
(hard dynamics \cite{MARR99}). In contrast, for factorizing (soft) dynamics 
the interface structure should remain independent of the field. This was 
recently confirmed for the unrestricted SOS interface 
by MC simulations \cite{RIKV01}. It is therefore important 
to use great caution in drawing conclusions about the microstructure of driven 
interfaces, based on dynamic MC simulations. For such conclusions to be valid, 
the dynamic must be chosen appropriately to the physical system of interest. 
The hard type of dynamics would appear to be particularly suited for certain 
interfaces in magnetic or dielectric systems, where the local order parameter is 
not conserved. 

One aspect of the interface dynamics not completely captured by our
model is, of course, the short-range correlations. Namely,
within the mean-field approximation used here, individual steps of the 
interface are assumed to be statistically independent. However, for increasing 
fields the interface undergoes a gradual breakdown of up/down symmetry. This is 
clearly seen in our simulations here, as well as in other examples of driven 
interfaces \cite{NEER97,KORN00}. It would seem likely that one could 
construct a mean-field approximation at the two-spin level, which might be able 
to predict this skewness for hard dynamics, as well as its absence for soft 
dynamics. However, such a theory has not yet been developed.

\section*{Acknowledgments}
\label{sec:ACK}

We acknowledge useful conversations with B.~Schmittmann and K.~Kawasaki. 
P.~A.~R.\ appreciates the hospitality of the 
Faculty of Integrated Human Studies at Kyoto University. 
The research was supported in part 
by National Science Foundation Grant Nos.~DMR-9981815 and DMR-0120310,
and by Florida State 
University through the Center for Materials Research and Technology and 
the School of Computational Science and Information Technology.



%
%
\begin{table}[ht]
\caption[]{
The spin classes in the anisotropic square-lattice SOS model. 
The first column contains the class labels, $jks$. 
The second column contains the total 
field and interaction energy for a spin in each class, $E(jks)$, relative 
to the energy of the state with all spins parallel and $H=0$, 
$E_0 = -2(J_x + J_y)$. 
The third column contains the change in the total system energy 
resulting from reversal of a spin from $s$ to $-s$, $\Delta E(jks)$. 
In both $E(jks)-E_0$ and $\Delta E(jks)$, the upper sign corresponds to $s=-1$,
and the lower sign to $s=+1$. 
The first three classes (marked  $\ast$) have nonzero populations in the 
SOS model, and flipping a spin in any of them preserves the SOS configuration. 
The other two classes also have nonzero populations in the 
SOS model, but flipping a spin in any of them would produce an overhang 
or a bubble and is therefore forbidden. 
}
\begin{tabular}{| l | l | l | }
\hline
Class, $jks$ 
& $E(jks) - E_0$ 
& $\Delta E(jks)$ 
\\ 
 \hline\hline
 $01s$ $\ast$
 & $\pm H + 2J_y$ 
 & $\mp 2H + 4J_x $
\\
 \hline
 $11s$  $\ast$
 & $\pm H + 2(J_x+J_y)$ 
 & $\mp 2H $
\\
 \hline
 $21s$  $\ast$
 & $\pm H + 2(2J_x+J_y)$ 
 & $\mp 2H - 4J_x $
\\
 \hline\hline
$10s$  
 & $\pm H + 2J_x$ 
 & $\mp 2H + 4J_y $
\\
 \hline
$20s$  
 & $\pm H + 4J_x$ 
 & $\mp 2H - 4(J_x-J_y) $
\\
\hline
 \end{tabular}
\label{table:class}
\end{table}

\begin{table}[ht]
\caption[]{
The mean populations for the spin classes of the SOS 
interface, with the corresponding contributions to the interface velocity 
under the hard 
Glauber dynamic. The first column contains the class labels, $jks$. 
The second column contains the mean spin-class populations for 
general tilt angle $\phi$, with $\cosh \gamma(\phi)$ from Eq.~(\ref{eq:chgam}). 
The third column contains the spin-class populations for $\phi = 0$. 
Using $X = e^{-2 \beta J_x}$ in these expressions yields the 
linear-response result in which the spin-class populations are 
evaluated for $H=0$. Using $X = X(T,H)$ from
Eq.~(\ref{eq:XTH}) with the transition probabilities of the
particular dynamic used [here: Glauber, Eq.~(\ref{eq:XG})], 
one gets the nonlinear-response approximation. 
The fourth column contains the contributions to the mean interface 
velocity in the $y$ direction from spins in classes $jk-$ and $jk+$, 
Eq.~(\protect\ref{eq_generalv}), using the SOS-preserving hard Glauber dynamic. 
}
\begin{tabular}{| l | l | l | l |}
\hline
Class, $jks$ 
& $\langle n(jks) \rangle$, general $\phi$ 
& $\langle n(jks) \rangle$, $\phi=0$ 
& $\langle v_y(jk) \rangle$ 
\\ 
 \hline\hline
 $01s$ $\ast$ 
 & $\frac{1 - 2X \cosh \gamma (\phi) + X^2}{(1-X^2)^2}$ 
 & $\frac{1}{(1+X)^2}$ 
 & $\frac{\tanh\left( \beta H \right)}
	 {1 + \left[\frac{\sinh \left(2 \beta J_x \right) }
		   {\cosh \left( \beta H \right)}\right]^2 }$  
\\
 \hline
 $11s$ $\ast$ 
 & $\frac{2X[(1+X^2) \cosh \gamma (\phi) - 2X]}{(1-X^2)^2}$ 
 & $\frac{2X}{(1+X)^2}$ 
 & ${\tanh\left( \beta H \right)}$
\\
 \hline
 $21s$ $\ast$ 
 & $\frac{X^2[1-2X\cosh\gamma(\phi)+X^2]}{(1-X^2)^2}$ 
 & $\frac{X^2}{(1+X)^2}$ 
 & $\frac{\tanh\left( \beta H \right)}
	 {1 + \left[ \frac{\sinh \left(2 \beta J_x \right) }
		   {\cosh \left( \beta H \right)}\right]^2 }$  
\\
 \hline\hline
$10s$ 
 & $\frac{2X^2}{1-X^2} 
\left\{
\frac{2\cosh^2\gamma(\phi)-1-2X\cosh\gamma(\phi)+X^2}
{1-2X\cosh\gamma(\phi)+X^2} \right.
$ 
 & $\frac{2X^2(1+2X)}{(1-X^2)(1+X)^2}$ 
 & 0  
\\
 &$ \left.
-
\frac{X^2 [1-2X\cosh\gamma(\phi)+X^2]}{(1-X^2)^2}
\right\}$ 
 &  
 & 
\\
 \hline
$20s$ 
 & $\frac{X^4 [1-2X\cosh\gamma(\phi)+X^2]}{(1-X^2)^3}$ 
 & $\frac{X^4}{(1-X^2)(1+X)^2}$ 
 & 0
\\
\hline
 \end{tabular}
\label{table:class2}
\end{table}
\clearpage 


\end{document}